\documentclass[aps,prd,showpacs,showkeys,floatfix,nofootinbib,twocolumn,
superscriptaddress,10pt]{revtex4-1}
\usepackage{natbib}
\usepackage{ulem,mwe}
\usepackage{color}
\usepackage{graphicx,verbatim}
\usepackage{epsfig}
\usepackage{bm}
\usepackage{amsthm}
\usepackage{amsfonts}
\usepackage{float}
\usepackage{amsmath,amssymb}
\usepackage{graphicx}
\usepackage[caption = false]{subfig}
\usepackage[colorlinks=true, citecolor=blue, urlcolor=blue]{hyperref}
\usepackage{cleveref}
\usepackage{xcolor}
\usepackage{comment}
\usepackage{tikz}
\usetikzlibrary{shapes,snakes}

\begin{document}
\title{Chemical freeze-out systematics of thermal model analysis using hadron yield ratios}
\author{Sumana Bhattacharyya}
\email{response2sumana91@gmail.com}
\affiliation{Department of Physics, Center for Astroparticle Physics \& Space Science, Bose Institute, Kolkata-700091, India}
\author{Amaresh Jaiswal}
\email{a.jaiswal@niser.ac.in}
\affiliation{School of Physical Sciences, National Institute of Science Education and Research, HBNI, Jatni-752050, Odisha, India}
\author{Sutanu Roy}
\email{sutanu@niser.ac.in}
\affiliation{School of Mathematical Sciences, National Institute of Science Education and Research, HBNI, Jatni-752050, Odisha, India}

%%%%%%%%%%%%%%%%%%%%%%%%%%%%%%%%%%%%%%%%%%%%%%%%%
\begin{abstract}
We provide a framework to estimate the systematic uncertainties in chemical freeze-out parameters extracted from $\chi^2$ analysis of thermal model, using hadron multiplicity ratios in relativistic heavy-ion collision experiments. Using a well known technique of graph theory, we construct all possible sets of independent ratios from available hadron yields and perform $\chi^2$ minimization on each set. We show that even for ten hadron yields, one obtains a large number ($10^8$) of independent sets which results in a distribution of extracted freeze-out parameters. We analyze these distributions and compare our results for chemical freeze-out parameters and associated systematic uncertainties with previous results available in the literature.
\end{abstract}
%%%%%%%%%%%%%%%%%%%%%%%%%%%%%%%%%%%%%%%%%%%%%%%%%


%12.38.Mh - Quark-gluon plasma
%21.65.Mn - Equations of state of nuclear matter
%24.10.Pa - Thermal and statistical models
%25.75.-q - Relativistic heavy-ion collisions

\maketitle

%%%%%%%%%%%%%%%%%%%%%%%%%%%%%%%%%%%%%%%%%%%%%%%%%

\section{Introduction}
\label{sec:Intro}

Relativistic heavy-ion collisions provides the opportunity to create hot and dense QCD matter and study its thermodynamic and transport properties. The `standard' model of relativistic heavy-ion collision has been developed in last few decades by analyzing the experimental data from Relativistic Heavy-Ion Collider (RHIC) at Brookhaven National Lab, USA and Large Hadron Collider (LHC)  at CERN in Geneva, Switzerland. The analysis of experimental data suggests that the fireball produced in these collisions consists of deconfined quarks and gluons in the early stage of evolution which tends to thermalize rapidly. This quark-gluon plasma (QGP) undergoes a transition from partonic phase to hadronic phase during the later stage of evolution and finally the energy-momentum of these hadrons are measured by the detectors. The rapid thermalization of the fireball, due to strongly interacting constituents of the medium, provides the motivation to look for thermal description of the observed hadron yields. Thermal models of hadronic system has a long history~\cite{Fermi:1950jd, Pomeranchuk:1951ey, Landau:1953gs, Belenkij:1956cd, Hagedorn:1980kb}. Indeed, statistical models, with the assumption of complete thermalization of hadronic matter, has been quite successful in explaining the hadron yields and their ratios measured in relatively recent experiments~\cite{Rischke:1991ke, Cleymans:1992jz, BraunMunzinger:1994xr, Cleymans:1996cd, Yen:1997rv, Heinz:1998st, Cleymans:1998fq, BraunMunzinger:1999qy, Cleymans:1999st, BraunMunzinger:2001ip, Becattini:2003wp, BraunMunzinger:2003zd, Karsch:2003zq, Tawfik:2004sw, Becattini:2005xt, Andronic:2005yp,Andronic:2008gu, Manninen:2008mg, Andronic:2012ut, Tiwari:2011km, Begun:2012rf, Fu:2013gga, Tawfik:2013eua, Garg:2013ata, Bhattacharyya:2013oya, Albright:2014gva, Kadam:2015xsa, Kadam:2015fza, Kadam:2015dda, Albright:2015uua, Bhattacharyya:2015zka, Bhattacharyya:2015pra, Begun:2016cva, Bhattacharyya:2017gwt, Andronic:2017pug, Ghosh:2018nqi, Dash:2018can}. 

Thermodynamic parameters obtained from statistical model analysis can be used to characterize the freeze-out hypersurface as the last surface of interaction. The onset of chemical freeze-out is said to have occurred when particle abundances are fixed and composition altering inelastic interactions have ceased. Subsequently kinetic freeze-out occurs when elastic interactions between particles have also stopped and the density drops to the level that the hadron momentum spectra remain unchanged. Beyond kinetic freeze-out, hadrons are assumed to stream freely to the detector. Hadron resonance gas (HRG) model, which assumes a statistical description of mixture of hadrons and their resonances in thermodynamic equilibrium, leads to a good description of the medium at freeze-out, for a wide range of collision energy~\cite{Cleymans:1998fq, Rischke:1991ke, Cleymans:1992jz, BraunMunzinger:1994xr, Cleymans:1996cd, Yen:1997rv, Heinz:1998st, BraunMunzinger:1999qy, Cleymans:1999st, BraunMunzinger:2001ip, Becattini:2003wp, BraunMunzinger:2003zd, Karsch:2003zq, Tawfik:2004sw, Becattini:2005xt, Andronic:2005yp,Andronic:2008gu, Manninen:2008mg, Andronic:2012ut, Tiwari:2011km, Begun:2012rf, Fu:2013gga, Tawfik:2013eua, Garg:2013ata, Bhattacharyya:2013oya, Albright:2014gva, Kadam:2015xsa, Kadam:2015fza, Kadam:2015dda, Albright:2015uua, Bhattacharyya:2015zka, Bhattacharyya:2015pra, Begun:2016cva, Bhattacharyya:2017gwt, Andronic:2017pug, Ghosh:2018nqi, Dash:2018can, Biswas:2020dsc, Biswas:2020kpu}. In this model, thermodynamic equilibrium state of the strongly interacting matter is completely determined by the temperature $(T)$ and the three chemical potentials $\mu_Q$, $\mu_B$ and $\mu_S$ corresponding to baryon number ($B$), electric charge ($Q$) and strangeness ($S$), respectively. These parameters at freeze-out can be extracted from statistical model calculations by performing a $\chi^2$ minimization fit to the available experimental multiplicity data~\cite{Andronic:2005yp, Andronic:2008gu, Andronic:2012ut, Alba:2014eba, Chatterjee:2015fua, Manninen:2008mg, Chatterjee:2017yhp, Adak:2016jtk, Cleymans:2004hj, Cleymans:2005xv, Chatterjee:2013yga}. Several codes like THERMUS~\cite{Wheaton:2004qb}, SHARE~\cite{Torrieri:2004zz}, THERMINATOR~\cite{Kisiel:2005hn} are publicly available to compute the abundances of particles using such statistical hadronization approach. 

The systematic uncertainties in the experimental data are expected to be reduced when one considers the ratios of hadron yields for $\chi^2$ analysis, which overall cancels the the system volume ($V$)~\cite{Andronic:2005yp, Adamczyk:2017iwn}. However, there are multiple ways to form a set of $(N-1)$ independent ratios given $N$ number of hadron yields. For simplicity, we call a set containing $(N-1)$ independent ratios as \textit{independent set}. It is evident that the choice of particle ratios, may introduce a bias for the extracted freeze-out parameters~\cite{Andronic:2005yp, Bhattacharyya:2019cer}. Therefore, the freeze-out parameters extracted from a $\chi^2$ minimization fit to these ratios would depend on the set under consideration. To avoid this problem, one may try to fit the absolute yields rather than their ratios which requires the inclusion of additional parameter $V$. This however is subject to its own bias~\cite{Andronic:2005yp, Andronic:2008gu, Andronic:2012ut}. It has also been observed that fitting absolute yield is more prone to converge to a false minima because of the strong correlation between yield normalization and other freeze-out parameters~\cite{Torrieri:2004zz}. This problem becomes more pronounced when one tries to incorporate excluded volume effect \cite{Andronic:2005yp}. On the other hand, yields ratios are not affected significantly when considering excluded volume effect and therefore the extracted freeze-out parameters remains stable.

In order to reliably extract freeze-out parameters from hadron yield ratios, one has to understand the systematic uncertainties due to the choice of specific ratios. First attempt towards understanding this uncertainty was made recently in Ref.~\cite{Bhattacharyya:2019cer} where the authors considered several independent sets for extraction of freeze-out parameters. These sets were chosen such that each of the $^NC_2$ ratios appear in at least one set. However it is important to note that the choice of these particular sets are also not unique. Therefore, the bias arising from choice of those specific sets still remains which could only be removed by considering all possible independent sets. Given the importance of statistical models in the context of relativistic heavy-ion collisions, it is essential to conclusively address the issue of systematic uncertainty.

In this article, we provide a framework to estimate systematic uncertainties in the extraction of chemical freeze-out parameters from analysis of hadron multiplicity ratios. We construct all possible independent sets from available hadron yields by using a well known technique in graph theory. Subsequently, we perform $\chi^2$ minimization on each set which leads to a distribution of the extracted freeze-out parameters. From these distributions, we obtain quantitative estimates of systematic uncertainty in the extracted freeze-out parameters corresponding to yield ratios of experimental data at $200$~GeV (RHIC) and $2.76$~TeV (LHC) collision energies. We also estimate these uncertainties after removing the usual constraints on the conserved charges. Finally, we compare our results for chemical freeze-out parameters and associated systematic uncertainties with previous results available in the literature.

%%%%%%%%%%%%%%%%%%%%%%%%%%%%%%%%%%%%%%%%%%%%%%%%%

\section{Hadron Resonance Gas model}
\label{sec:hrg}

We consider the HRG model for our analysis of yield ratios. The thermodynamic potential of HRG in terms of grand-canonical partition function is given by,
\begin{equation}
\ln Z^{\rm id}=\pm\!\sum_i\! \frac{Vg_i}{(2\pi)^3}\!\int\! d^3p \,\ln\!\left[1\pm\exp\!\left(\!-\frac{E_i-\mu_i}{T}\!\right)\!\right]\!,
\end{equation}
where sum runs over all hadrons and resonances. The upper sign is for fermions and lower is for bosons, and the normalization factor $V$ is the volume of the fireball. Here $g_i$, $E_i$ and $m_i$ are respectively the degeneracy factor, energy and mass of $i^{th}$ hadron. While, $\mu_i=B_i\mu_B+S_i\mu_S+Q_i\mu_Q$ is the chemical potential of the $i^{th}$ hadron with $B_i$, $S_i$ and $Q_i$ denoting its baryon number, strangeness and electric charge.

For a thermalized system the number density $n_i$ obtained from partition function is given as,
\begin{equation}
n_i =\frac{T}{V}\!\left(\!\frac{\partial \ln Z_i}{\partial\mu_i}\!\right)_{\!\!V,T} \!\!\!=\frac{g_i}{{(2\pi)}^3}\!\int\!\!\frac{d^3p} {\exp[(E_i-\mu_i)/T]\pm1}.
\end{equation}
The rapidity density for $i$'th detected hadron to the corresponding number density in the HRG model can be written as 
~\cite{Manninen:2008mg},
\begin{equation}
\frac{dN_i}{dy}\Biggr|_{\rm Det} \simeq {\frac{dV}{dy}}n_i^{\rm Tot}\Biggr|_{\rm Det}
\end{equation}
where the subscript `Det' denotes the detected hadrons. If the heavier resonances $j$ decay to the $i$-th hadron, then 
\begin{equation}
n_i^{\rm Tot} = n_i 
+\sum_{j} n_j \times {\rm branching~ratio~of}\,(j\rightarrow i),
\end{equation}
where $n_i$ and $n_j$ depend on the thermal parameters $(T,\mu_B,\mu_Q,\mu_S)$  which can be obtained by fitting experimental hadron yields to the model calculations. The fitted values of thermal parameters are obtained by minimizing the $\chi^{2}$, which is defined as,
\begin{equation}
\label{eq:chisqr}
\chi^{2}=\sum_{\alpha} \frac{(R_{\alpha}^{\rm exp}-R_{\alpha}^{\rm th})^{2}}
{(\sigma^{\rm exp}_{\alpha})^2},
\end{equation}
where, $R_{\alpha}^{\rm exp}$ is the ratio of hadron yields obtained from experiments, $R_{\alpha}^{\rm th}$ is the ratio of the number densities from the model calculations and $\sigma^{\rm exp}_{\alpha}$ is the experimental uncertainty. 

As individual multiplicities or their ratios are not conserved by strong interactions, $\chi^2$ fitting is performed by choosing a set of multiplicity ratios. In general, with $N$ given experimental hadron yields $p_{1}, p_{2}, \cdots ,p_{N}$ one can construct $N(N-1)$ ratios of the form $p_{i}/p_{j}$, where $1\leq i,j \leq N$ and $i\neq j$. However, interchanging numerator and denominator does not provide any new information in our analysis. Therefore, it is sufficient to consider $N(N-1)/2$ ratios of the given particle yields. Because of the correlation existed between all possible multiplicity ratios, it is only relevant to choose $(N-1)$ number of statistically independent ratios out of those $N(N-1)/2$ ratios to parameterize the thermal model~\cite{Andronic:2005yp, Becattini:2007wt}. It has also been pointed out within a thermal model analysis that, choosing a specific independent set may introduce bias in the minimization process~\cite{Bhattacharyya:2019cer}. However, it is possible to quantify this systematic uncertainty in the freeze-out parameters, extracted from $\chi^2$ fitting, by considering \textit{all possible sets of independent ratios}. Observe that, $(N-1)$ ratios can be chosen in $^{N(N-1)/2}C_{N-1}$ ways. But, any set containing $(N-1)$ ratios may not necessarily be independent. Also, enumerating all independent sets from $^{N(N-1)/2}C_{N-1}$ sets, is a complex and tedious process because the total number of independent sets grows immensely for large $N$. We perform this enumeration process by relating it to a particular case of the \textit{minimum spanning tree enumeration problem}, one of the well-known problems in combinatorial optimization.

%%%%%%%%%%%%%%%%%%%%%%%%%%%%%%%%%%%%%%%%%%%%%%%%%

\section{Generating all independent sets}
\label{sec:sets}

Let~\(p_{1}, p_{2},\cdots , p_{N}\) be \(N\) distinct particle yields. Assume~\(S\) be an \textit{independent set} containing the ratios of these particle yields denoted as $r_{1}, r_{2}, \cdots ,r_{N-1}$. We associate an undirected graph $G$ to $S$ whose vertices are labeled by $p_{1}, p_{2}, \cdots ,p_{N}$ and there is an edge $\overline{p_{i}p_{j}}$ between the vertices $p_{i}, p_{j}$, if either $p_{i}/p_{j}$ or $p_{j}/p_{i}$ is in $S$. Observe that if $\overline{p_{i}p_{j}}, \overline{p_{j}p_{k}}$ are edges of $G$ then the vertices $p_{i}$ and $p_{k}$ are connected through the vertex $p_{j}$. More generally, a path between two vertices $p_{i}$ and $p_{j}$ is said to exists if either there is an edge between them or they can be connected through some or all of the remaining $N-2$ vertices. Since each of the particle yields $p_{1}, p_{2}, \cdots ,p_{N}$ must appear at least in one of the ratios in $S$, then every vertex of $G$ is connected to at least one of the remaining $N-1$ vertices. Independence of $S$ also implies that for each $1\leq k\leq N-1$
\begin{equation}\label{eq:indset}
r_{k}\neq \displaystyle\prod_{\substack{l=1\\ l\neq k}}^{N-1} r_{l}^{s_{l}} \quad \text{where  $s_{l}=0,\pm 1$.}
\end{equation}
The above equation imposes the condition that if $\overline{p_{i}p_{j}}$ is an edge of $G$ then there is no other path between $p_{i}$ and $p_{j}$. Equivalently, between any two vertices of $G$ there is a unique path between them. Hence, $G$ is a \textit{tree} on the $N$ vertices labeled by $p_{1}, p_{2}, \cdots ,p_{N}$. Conversely, every tree on $N$ vertices $p_{1}, p_{2}, \cdots ,p_{N}$ gives rise to an independent set. Therefore, we have a one to one correspondence between the trees on $N$ vertices labeled by $p_{1}, p_{2}, \cdots ,p_{N}$ and the independent sets. 

\begin{figure}[t]
\includegraphics[width=\linewidth]{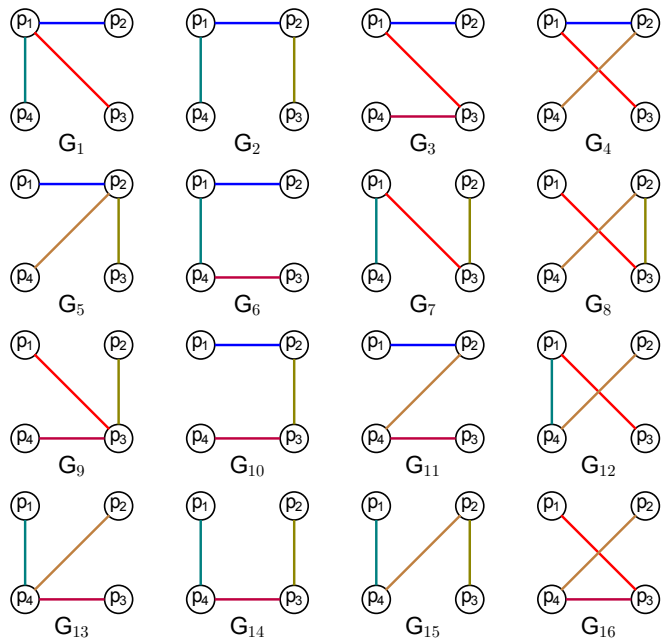}
\caption{All possible tree diagrams for $N=4$ case. The vertices symbolize particle yields. Each diagram corresponds to one set of independent ratios where the ratios are represented by lines joining two vertices. There are six distinct ratios (unique upto inverse) which are denoted by different colors.}
\label{tree}
\end{figure}

For example, let us consider four distinct particles \(p_{1}, p_{2}, p_{3}, p_{4}\). Then we have the following \(16\) independent sets:
\begin{alignat*}{2}
I_{1} &=\left\{\frac{p_2}{p_1},\frac{p_3}{p_1},\frac{p_4}{p_1}\right\},\quad 
I_{2} &=\left\{\frac{p_2}{p_1},\frac{p_4}{p_1},\frac{p_3}{p_2}\right\},\\ 
I_{3} &=\left\{\frac{p_2}{p_1},\frac{p_3}{p_1},\frac{p_4}{p_3}\right\},\quad
I_{4} &=\left\{\frac{p_2}{p_1},\frac{p_3}{p_1},\frac{p_4}{p_2}\right\},\\ 
I_{5} &=\left\{\frac{p_2}{p_1},\frac{p_3}{p_2},\frac{p_4}{p_2}\right\},\quad 
I_{6} &=\left\{\frac{p_2}{p_1},\frac{p_4}{p_1},\frac{p_4}{p_3}\right\},\\
I_{7} &=\left\{\frac{p_3}{p_1},\frac{p_4}{p_1},\frac{p_3}{p_2}\right\},\quad
I_{8} &=\left\{\frac{p_3}{p_1},\frac{p_3}{p_2},\frac{p_4}{p_2}\right\},\\
I_{9} &=\left\{\frac{p_3}{p_1}, \frac{p_3}{p_2},\frac{p_4}{p_3}\right\},\quad 
I_{10} &=\left\{\frac{p_2}{p_1},\frac{p_3}{p_2},\frac{p_4}{p_3}\right\},\\
I_{11} &=\left\{\frac{p_2}{p_1},\frac{p_4}{p_2},\frac{p_4}{p_3}\right\},\quad 
I_{12} &=\left\{\frac{p_3}{p_1},\frac{p_4}{p_1},\frac{p_4}{p_2}\right\},\\
I_{13} &=\left\{\frac{p_4}{p_1},\frac{p_4}{p_2},\frac{p_4}{p_3}\right\},\quad 
I_{14} &=\left\{\frac{p_4}{p_1},\frac{p_3}{p_2},\frac{p_4}{p_3}\right\},\\
I_{15} &=\left\{\frac{p_4}{p_1},\frac{p_3}{p_2},\frac{p_4}{p_2}\right\},\quad 
I_{16} &=\left\{\frac{p_3}{p_1},\frac{p_4}{p_2},\frac{p_4}{p_3}\right\}.
\end{alignat*}
These sets can be obtained by associating a unique tree~\(G_{m}\) to every independent set~\(I_{m}\) for all~\(1\leq m\leq 16\) as shown in Fig.~\ref{tree}. Here the vertices symbolize particle yields. Each diagram in Fig.~\ref{tree} corresponds to one set of independent ratios where the ratios are represented by lines joining two vertices. There are six distinct ratios (unique up to inverse) for $N=4$ which are denoted by different colors. In general, the total number of different trees on~\(N\) labeled vertices is equal to $N^{N-2}$, which is given by the Cayley's formula~\cite{Cayley:1889, Cayley:1897}. In fact the generation of those $N^{N-2}$ many trees is equivalent to generating all spanning trees of the complete graph on $N$ vertices labeled by $p_{1}, p_{2}, \cdots , p_{N}$. This is a particular case of the minimum spanning tree enumeration problem. This problem is important in its own right due to its wide range of applications including telecommunication networks and therefore several algorithms with improvised efficiencies have been developed since mid 50's. We refer~\cite{Graham1985, Cormen2009, Chakraborty:2019algo} and the references therein for more details.

%%%%%%%%%%%%%%%%%%%%%%%%
%%%%%%%%%%%%%%%%%%%%%%%%

The algorithm we have employed to generate all trees on \(N\) labeled vertices is as follows. It is straightforward to check that there can be~\(\frac{N(N-1)}{2}\) different edges connecting any two vertices~\(p_{i}, p_{j}\) considering ~\(1\leq i,j\leq N\) and~\(i\neq j\). By implementing the algorithm, at every step we choose~\(N-1\) edges say~\(e_{1}, e_{2}, \cdots , e_{N-1}\) (without repetition) from~\(\frac{N(N-1)}{2}\) edges and denote the graph formed with $N$ vertices labeled by~\(p_{1}, p_{2}, \cdots , p_{N}\) and edges~\(e_{1}, e_{2}, \cdots ,e_{N-1}\), by~$G'$. In the next step, we shall verify whether the vertex~\(p_{i}\), for all~\(1\leq i\leq N\), is connected to any other vertex by an edge or not. Then we construct the incidence matrix~\(M=(m_{ij})_{N\times (N-1)}\) to verify whether the graph~\(G'\) is a tree. The entries of~\(M\) are defined as follows:~\(m_{ij}=1\) whenever the edge~\(e_{j}\) joins the vertex~\(p_{i}\) to some other vertex, and~\(0\) otherwise, for all~\(1\leq i,j\leq N\). Observe that, if we delete one of the rows of~\(M\) then we obtain a submatrix of~\(M\) of size~\((N-1)\times (N-1)\). We continue this process for each of the~\(N\) rows of~\(M\) and obtain~\(N\) many submatrices of~\(M\). Then we check whether the rank of each of those $N$ submatrices of~\(M\) is~$N-1$. If it is true then we conclude that~$G'$ is a tree. In this way, we generate all~\(N^{N-2}\) trees on~\(N\) vertices and consequently, we obtain all~\(N^{N-2}\) possible independent sets of ratios for our analysis.

%%%%%%%%%%%%%%%%%%%%%%%%%%%%%%%%%%%%%%%%%%%%%%%%%

\section{Data analysis}
\label{sec:analysis}

We have used RHIC~\cite{Kumar:2012fb, Das:2012yq, Adler:2002uv, Adams:2003fy, Zhu:2012ph, Zhao:2014mva, Kumar:2014tca, Das:2014kja, Abelev:2008ab, Aggarwal:2010ig, Abelev:2008aa, Adcox:2002au, Adams:2003fy, Adler:2002xv, Adams:2006ke, Adams:2004ux, Kumar:2012np, Adams:2006ke} and LHC~\cite{Abelev:2012wca, Abelev:2013xaa, ABELEV:2013zaa, Abelev:2013vea} data at mid-rapidity for the most central collisions for our analysis. All hadrons with masses up to $2$~GeV, with known degrees of freedom, are included for the HRG spectrum. The masses and branching ratios used are as given in Refs.~\cite{Wheaton:2004qb, Tanabashi:2018oca}. It is to be noted that we have used vacuum masses for all the hadrons as the in-medium mass modification in connection with chiral symmetry restoration does not have any significant effect on freeze-out parameters for higher collision energies~\cite{Zschiesche:2002zr, Koch:2003pj, Brown:2001nh}. We could not find the $\bar{\Lambda}$ yield at LHC. Therefore at this energy, we assumed $\bar{\Lambda}$ yield to be same as that reported for ${\Lambda}$. Therefore, in order to obtain the freeze-out parameters, we have used the hadron yields of $\pi^\pm$ (139.57 MeV), $k^{\pm}$ ($493.68$~MeV), $p$,$\bar{p}$ ($938.27$~MeV), $\Lambda$,$\bar{\Lambda}$ ($1115.68$~MeV) and $\Xi^\mp$ ($1321.71$~MeV).

To obtain the freeze-out parameters it is a common approach to fix $\mu_Q$ and $\mu_S$ using these following constraint relations \cite{Alba:2014eba} , 
\begin{equation}\label{eq:nbq}
\frac{\sum_i n_i (T, \mu_B, \mu_S, \mu_Q) B_i}{ \sum_i n_i (T, \mu_B, 
\mu_S,\mu_Q) Q_i} = {\rm constant},
\end{equation}
and
\begin{equation}\label{eq:ns}
\sum_i n_i (T, \mu_B, \mu_S, \mu_Q) S_i=0.
\end{equation}
With these two constraint equations, the problem is reduced to a two dimensional problem. The constant value of the ratio of net baryon number and net charge, depends on the physical system. For example, in Au + Au collisions, this constant is given by $(N_p + N_n)/N_p = 2.5$, with $N_p$ and $N_n$ denoting the number of protons and neutrons in the colliding nuclei. However for LHC ($\sqrt{s_{NN}}=2.76$~TeV) and RHIC ($\sqrt{s_{NN}}=200$~GeV) energies, baryon stopping is expected to be negligible. In such cases, net baryon number and net charge may be vanishingly small and therefore their ratio may become arbitrary in a rapidity bin. Thus, it seems natural to relax the constraint given in Eq.~\eqref{eq:nbq} to extract freeze-out parameters by $\chi^2$ minimization. In the present analysis, we consider both cases: we call \textbf{Model-I} where the constraint given in Eq.~\eqref{eq:nbq} is imposed and \textbf{Model-II} where we do not enforce this constraint relation to hold.

For $N=10$ observed hadron yields, we have first obtained $10^8$ independent sets of nine ratios, as explained in Sec.~\ref{sec:sets}. We then performed $\chi^2$ minimization fits for each set using our numerical code for both Model-I and Model-II with a convergence criteria of $10^{-4}$ or better. This procedure to fit all possible independent sets results in $10^8$ values of each extracted freeze-out parameters and corresponding $\chi^2$ numbers. We then construct histograms for each extracted freeze-out parameters, as well as for the $\chi^2$ values. However, histograms give an overall notion about the density of the underlying distribution of the data. Therefore, for further analysis, we perform a least square fit to histograms of each parameter with Gaussian distribution. %and the histogram of $\chi^2$, with $\chi^2$ distribution. 
The fitted distribution curves quantify the probability distribution function for a population that has the maximum likelihood of producing the distribution that exists in the given sample. The mean of the Gaussian distribution corresponding to a given freeze-out parameter leads to an estimate of the central value of that parameter. The systematic uncertainties arising from the fitting procedure is calculated from full width at half maximum $(2.3548\sigma)$, where $\sigma$ is the standard deviation of the Gaussian distribution.

%%%%%%%%%%%%%%%%%%%%%%%%%%%%%%%%%%%%%%%%%%%%%%%%%

\section{Results and discussions}
\label{sec:result}

In this section, we discuss the extracted chemical freeze-out parameters for center of mass collision energy $\sqrt{s_{NN}}=200$~GeV at RHIC and $\sqrt{s_{NN}}=2.76$~TeV at LHC. We fit $10^8$ sets of yield ratios to obtain a distribution for fit parameters, i.e., freeze-out temperatures $T$ and freeze-out chemical potentials $\mu_B$, $\mu_Q$ and $\mu_S$. The estimated value of the parameters are derived from the mean of the fitted Gaussian distribution. The error bars quantifies the variations of the parameters obtained from all possible sets of independent ratios. Results including the constraint, Eq.\eqref{eq:nbq}, on net baryon and net charge (Model-I) and those for the unconstrained system (Model-II), are shown separately. Finally we also compare our results with those obtained earlier in literature.

%%%%%%%%%%%%%%%%%%%%%%%%
\begin{figure*}[!htb]
\centering
\subfloat[]{
{\includegraphics[scale=0.7]{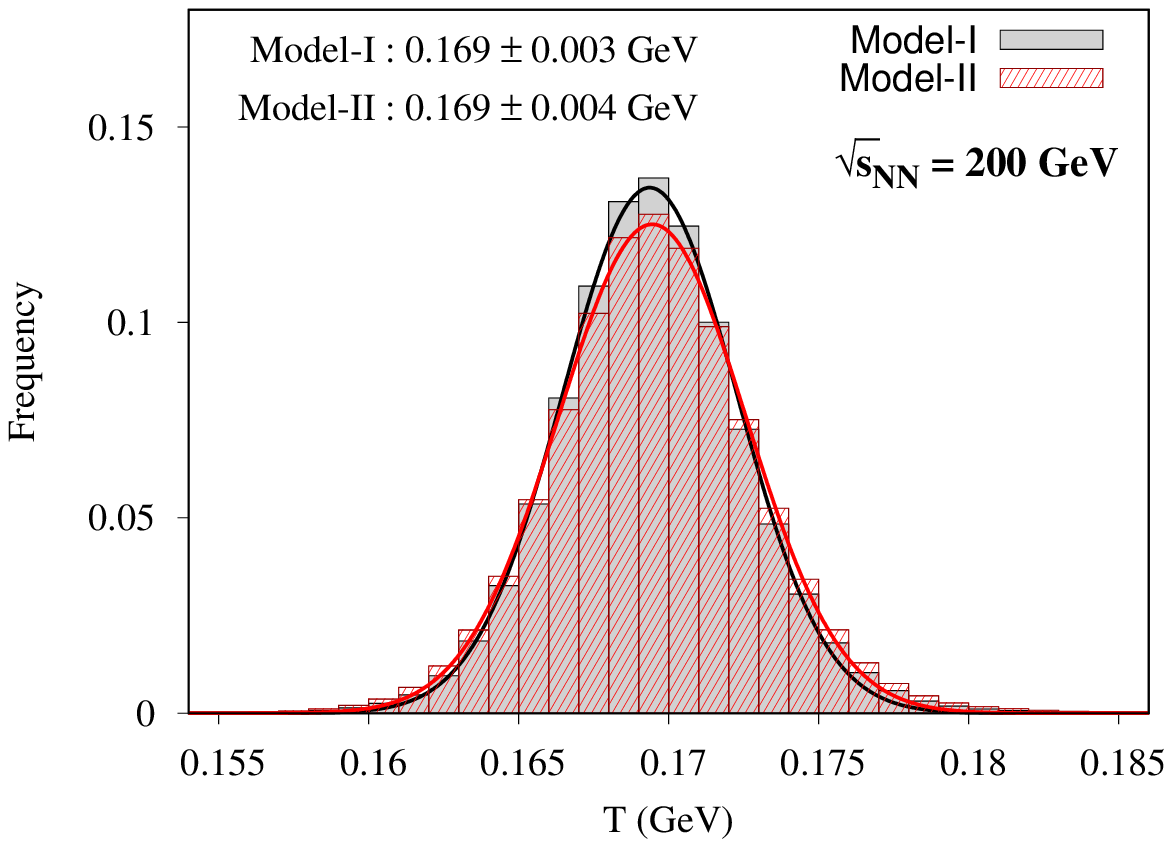}}
\label{fg.TRHIC}}
\subfloat[]{
{\includegraphics[scale=0.7]{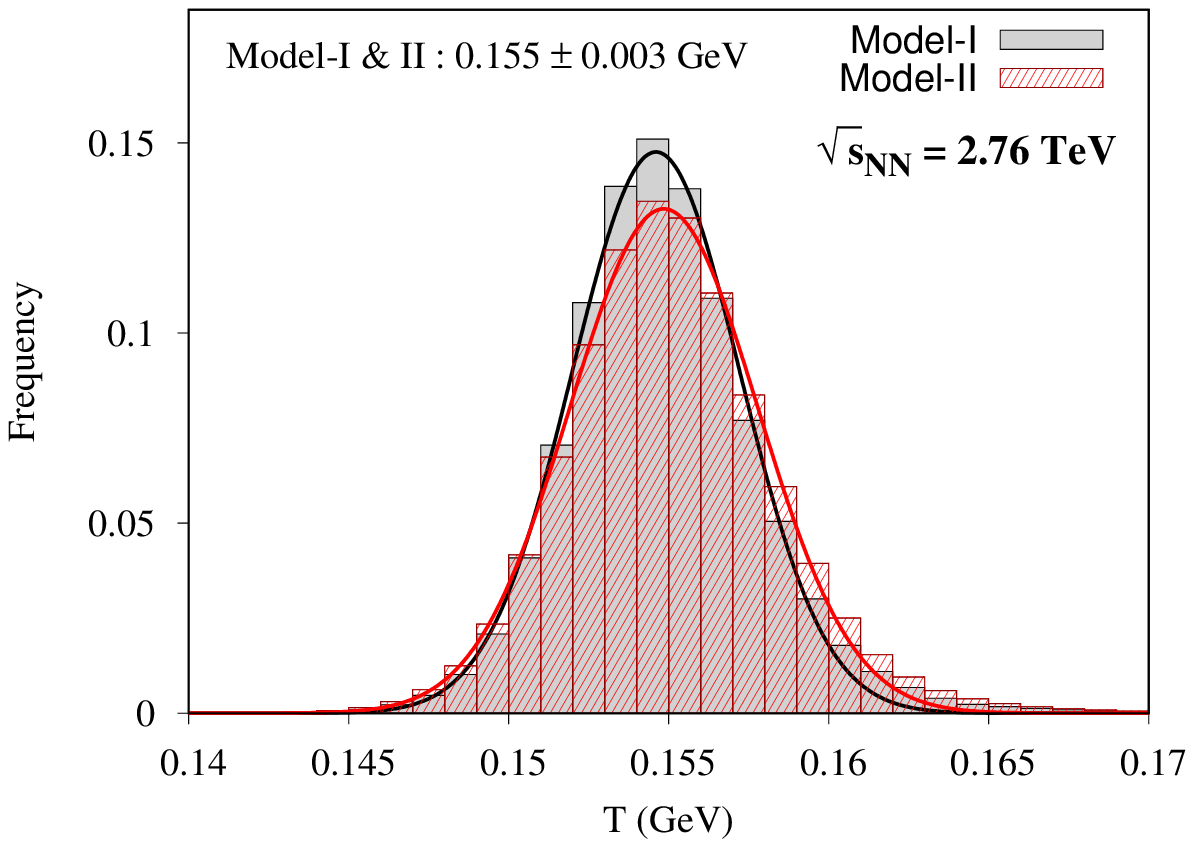}}
\label{fg.TLHC}}\\
\caption{The distribution of chemical freeze-out temperatures for collision energies $\sqrt{s_{NN}}=200$~GeV (RHIC) and $\sqrt{s_{NN}}=2.76$~TeV (LHC). Also shown are the fitted Gaussian curve and the values of extracted freeze-out temperature with systematic error bars are listed, for both Model-I and Model-II.}
\label{fg.T}
\end{figure*}
%%%%%%%%%%%%%%%%%%%%%%%%

%%%%%%%%%%%%%%%%%%%%%%%%
\begin{figure*}[!htb]
\centering
\subfloat[]{
{\includegraphics[scale=0.7]{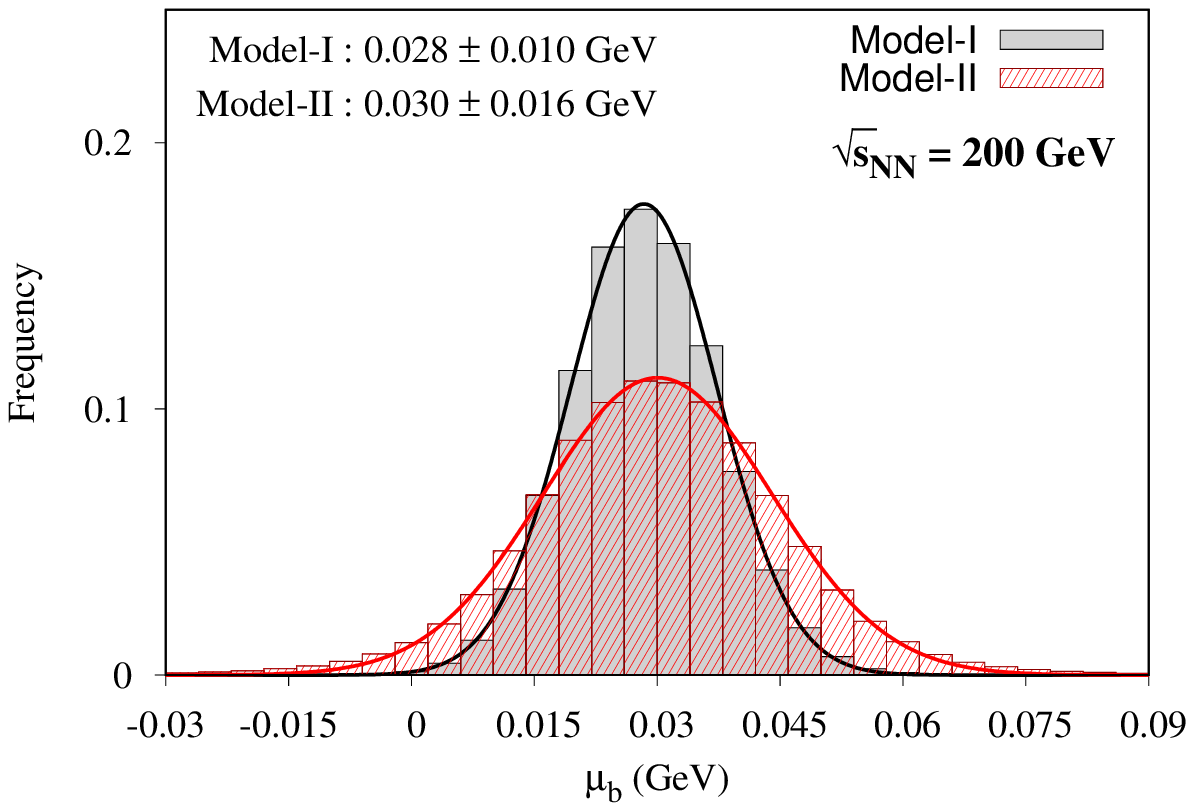}}
\label{fg.mubRHIC}}
\subfloat[]{
{\includegraphics[scale=0.7]{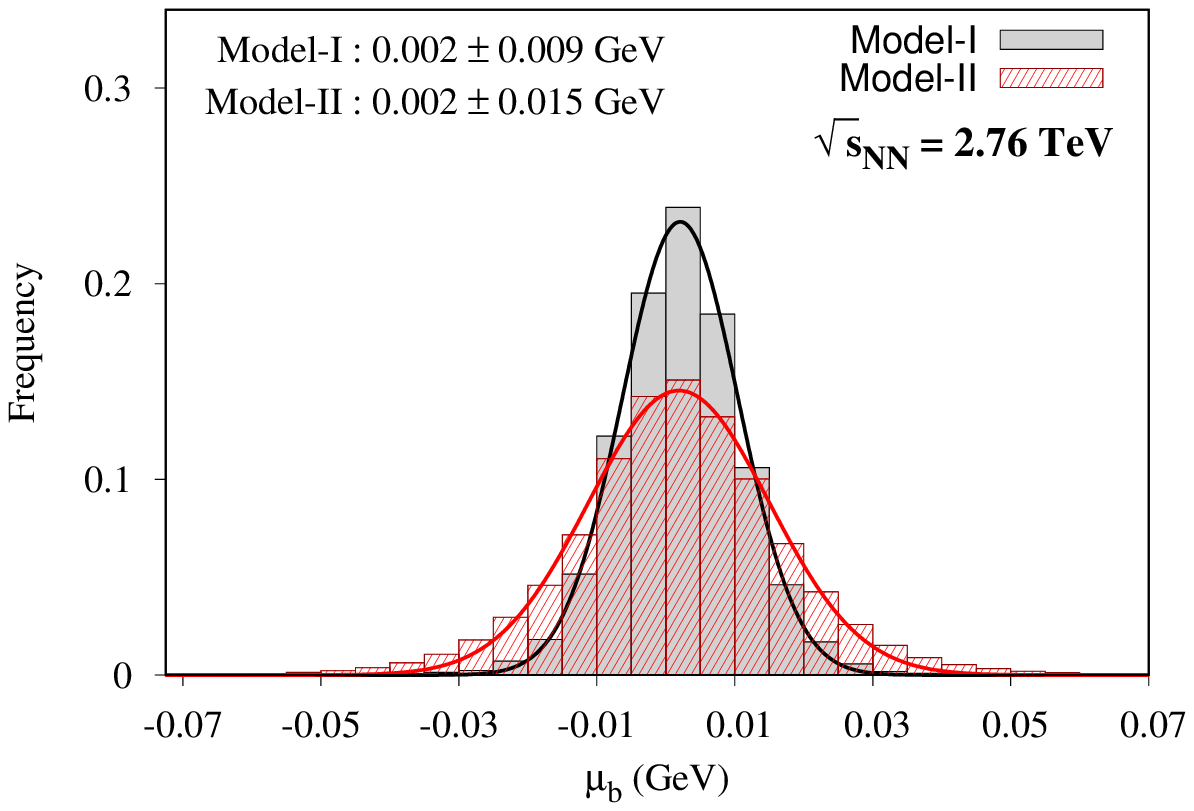}}
\label{fg.mubLHC}}\\
\caption{
\label{fg.mub}
The distribution of baryon chemical potential at chemical freeze-out for collision energies $\sqrt{s_{NN}}=200$~GeV (RHIC) and $\sqrt{s_{NN}}=2.76$~TeV (LHC). Also shown are the fitted Gaussian curve and the values of extracted baryon chemical potential freeze-out with systematic error bars are listed, for both Model-I and Model-II.}
\end{figure*}
%%%%%%%%%%%%%%%%%%%%%%%%

Histogram distribution and corresponding Gaussian fit to this distribution for freeze-out temperatures are shown for center of mass energy $\sqrt{s_{NN}}=200$~GeV in Fig.~\ref{fg.TRHIC} and $\sqrt{s_{NN}}=2.76$~TeV in Fig.~\ref{fg.TLHC}. The qualitative as well as the quantitative estimations, for both Model-I and Model-II, are consistent with each other. We find that for RHIC the extracted mean value of freeze-out temperature is $169$~MeV which is slightly larger than the earlier estimates. On the other hand, for LHC, we find the fitted mean value to be $155$~MeV. Nevertheless the estimated values lies in near proximity with the values reported in literature, where the analysis was done with a particular set of independent ratios as documented in Table~\ref{tb.RHIC} and in Table~\ref{tb.LHC}. Our analysis shows that the present model has a small systematic uncertainty in chemical freeze-out temperature, which is quite promising. In general, the freeze-out temperature increases with increasing $\sqrt{s_{NN}}$ and approaches a saturation~\cite{Hagedorn:1965st}, except at the LHC energy. As expected, in our analysis also, the temperature is lower at LHC energy than RHIC energy, due to the lower yield of protons at LHC~\cite{Abelev:2013vea}.

In Fig.~\ref{fg.mub}, we show the histogram distribution and the Gaussian fit to this distribution for baryon chemical potential $\mu_B$ extracted from all independent sets. In Fig.~\ref{fg.mubRHIC} we present our results for $\sqrt{s_{NN}}=200$~GeV collision energy whereas in Fig.~\ref{fg.mubLHC}, results for $\sqrt{s_{NN}}=2.76$~TeV is shown. Generally, for lower collision energies, significant number of baryons deposit their energies in the vicinity of center of mass frame due to baryon stopping. But as the energy increases colliding baryons may become almost transparent to each other and deposit their energy outside the collision region. In our analysis for both the models, we have estimated the value of $\mu_B$ and quantifies the systematic uncertainty for high collision energies. The comparison with several literature show our derived values lie in the range of their predicted values, see Table~\ref{tb.RHIC} and~\ref{tb.LHC} for reference. 

%%%%%%%%%%%%%%%%%%%%%%%%
\begin{figure*}[!htb]
\centering
\subfloat[]{
{\includegraphics[scale=0.7]{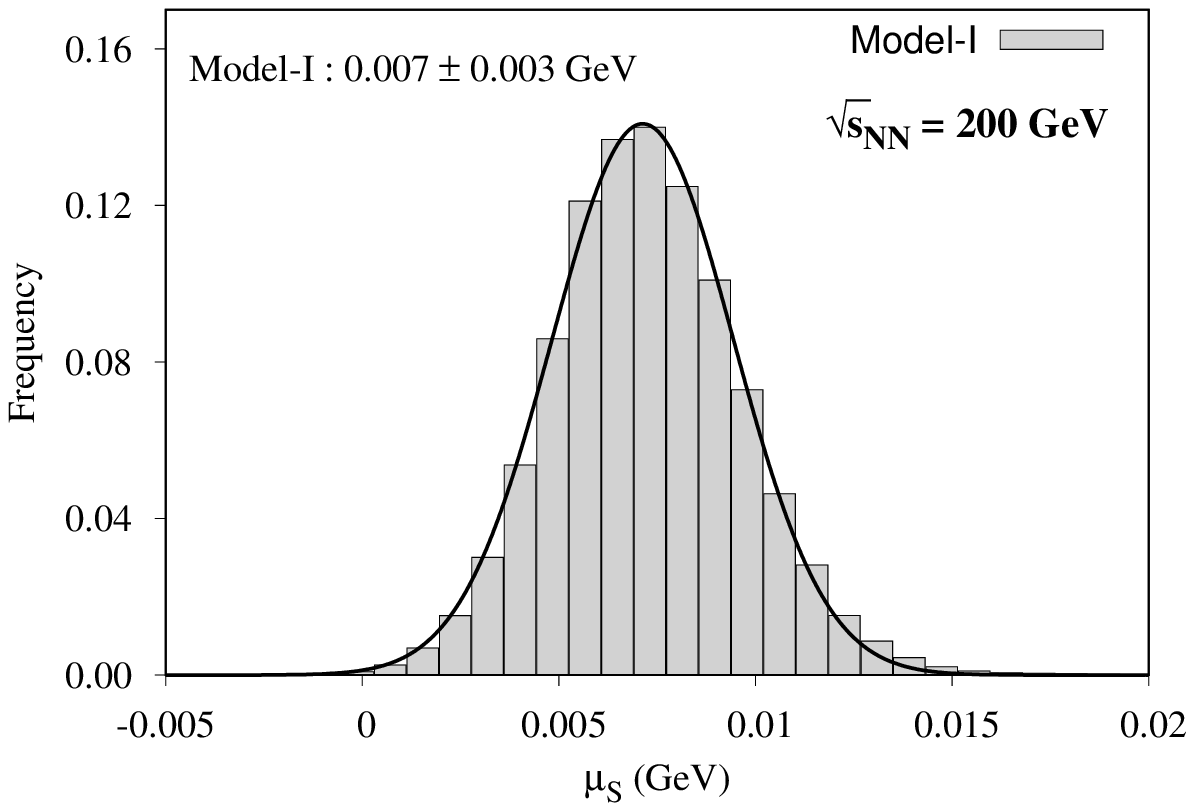}}
\label{fg.musRHICwc}}
\subfloat[]{
{\includegraphics[scale=0.7]{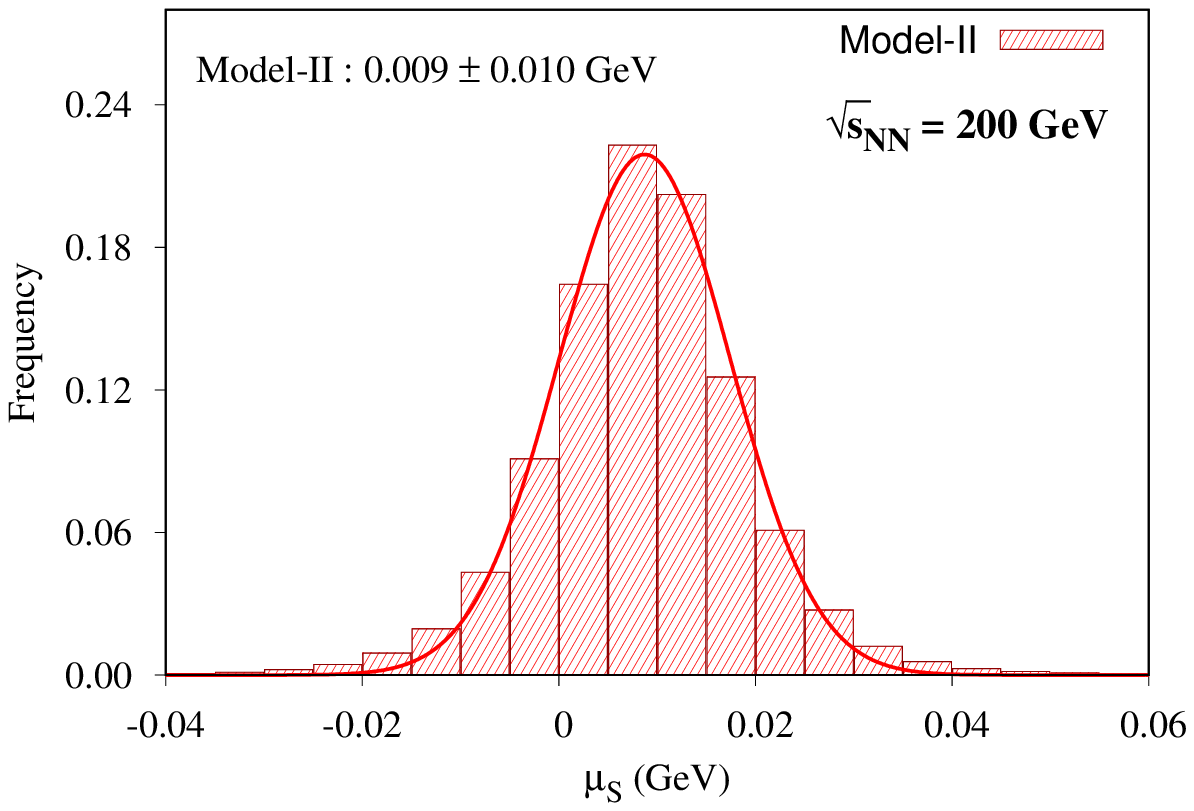}}
\label{fg.musRHICwoc}}\\
\subfloat[]{
{\includegraphics[scale=0.7]{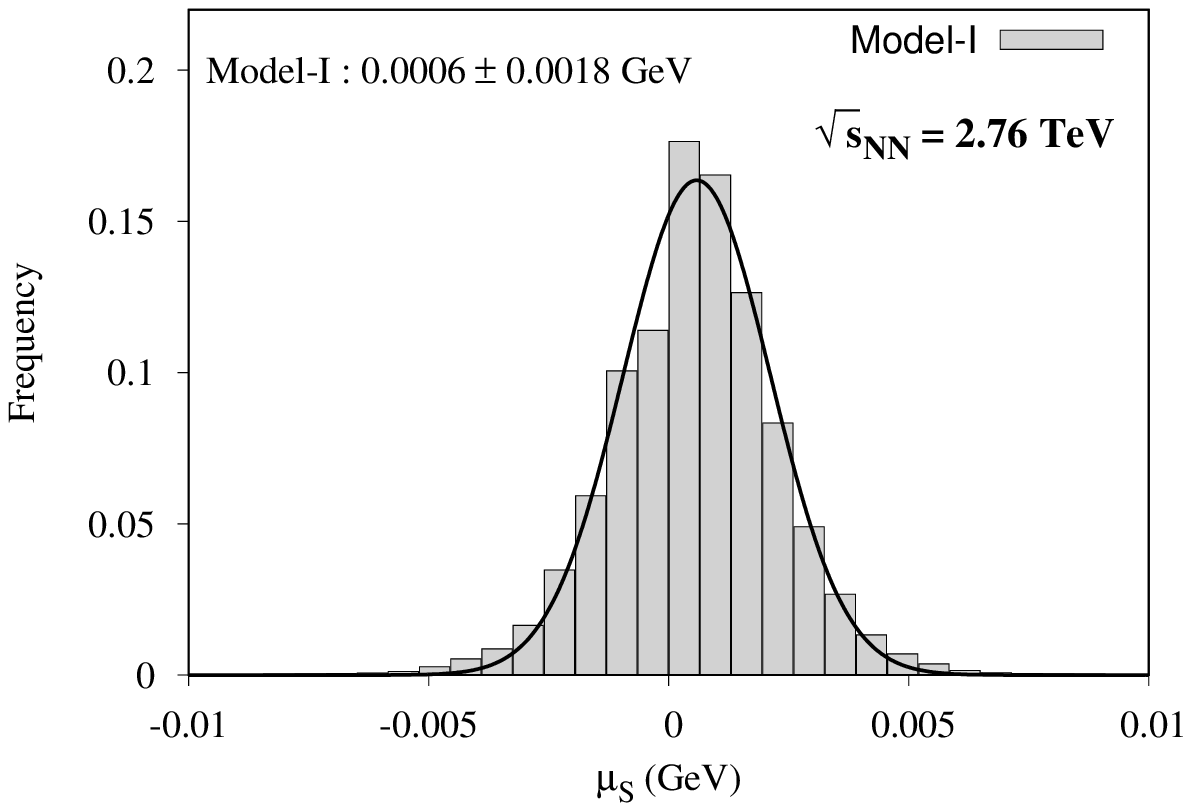}}
\label{fg.musLHCwc}}
\subfloat[]{
{\includegraphics[scale=0.7]{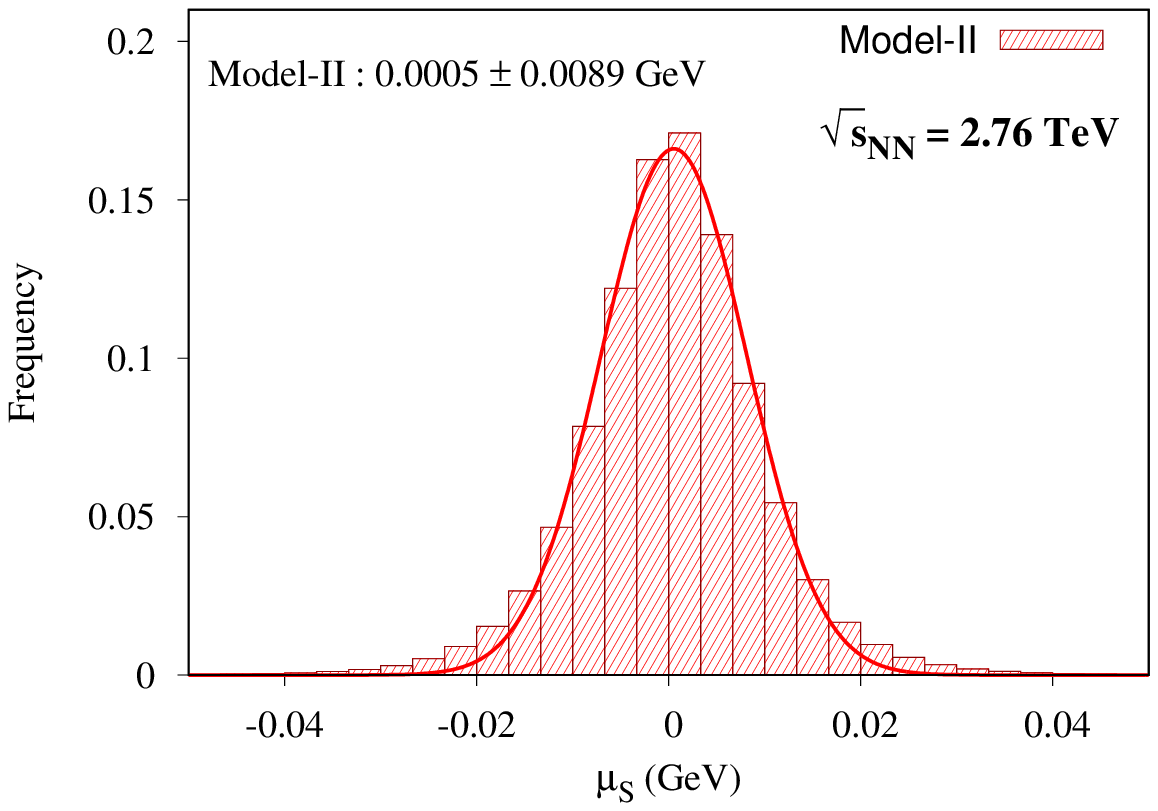}}
\label{fg.musLHCwoc}}\\
\caption{
\label{fg.musmuqRHIC}
The distribution of strange chemical potential at chemical freeze-out for collision energies $\sqrt{s_{NN}}=200$~GeV (RHIC) and $\sqrt{s_{NN}}=2.76$~TeV (LHC). Also shown are the fitted Gaussian curve and the values of extracted strange chemical potential at freeze-out with systematic error bars are listed, for both Model-I and Model-II.}
\end{figure*}
%%%%%%%%%%%%%%%%%%%%%%%%

Due to the possible redistribution of Fermi momentum among larger degrees of freedom, strangeness production at higher baryon densities is conjectured to be dominant. The non-zero value of baryon chemical potential induces the production of strange baryons, which imminently requires the strange chemical potential to produce enough strange anti-particles. The existence of this non-zero strange chemical potential $\mu_S$, ensures the vanishing of net strangeness condition of the colliding nuclei. The appearance of non-zero $\mu_S$ is considered as an indication of chemical equilibration in strongly interacting matter~\cite{Witten:1984rs}. As the collision energy increases, $\mu_S$ eventually decreases. On the other hand, electric charge chemical potential $\mu_Q$ appears due to the net-isospin of the colliding nuclei. While scanning the collision energy range, $\mu_Q$ remains non-zero but quite small and eventually approaches to zero at higher $\sqrt{s_{NN}}$. Several analyses thus approximated $\mu_Q$ to be zero at higher collision energies~\citep{Adamczyk:2017iwn}. In our analysis, we have estimated the value of $\mu_S$ and $\mu_Q$ using both Model-I and Model-II. 

The mean value of $\mu_S$'s are in reasonable agreement with each other for both the models, as illustrated in Figs.~\ref{fg.musRHICwc}~and~\ref{fg.musRHICwoc} for RHIC energy. Though the estimated values commensurate for both the models, Model-II has much larger systematic uncertainty and exclusion of constraint is the source of this uncertainty. The same features are also observed for LHC energy as shown in Figs.~\ref{fg.musLHCwc}~and~\ref{fg.musLHCwoc}. 

%%%%%%%%%%%%%%%%%%%%%%%%
\begin{figure*}[!htb]
\centering
\subfloat[]{
{\includegraphics[scale=0.7]{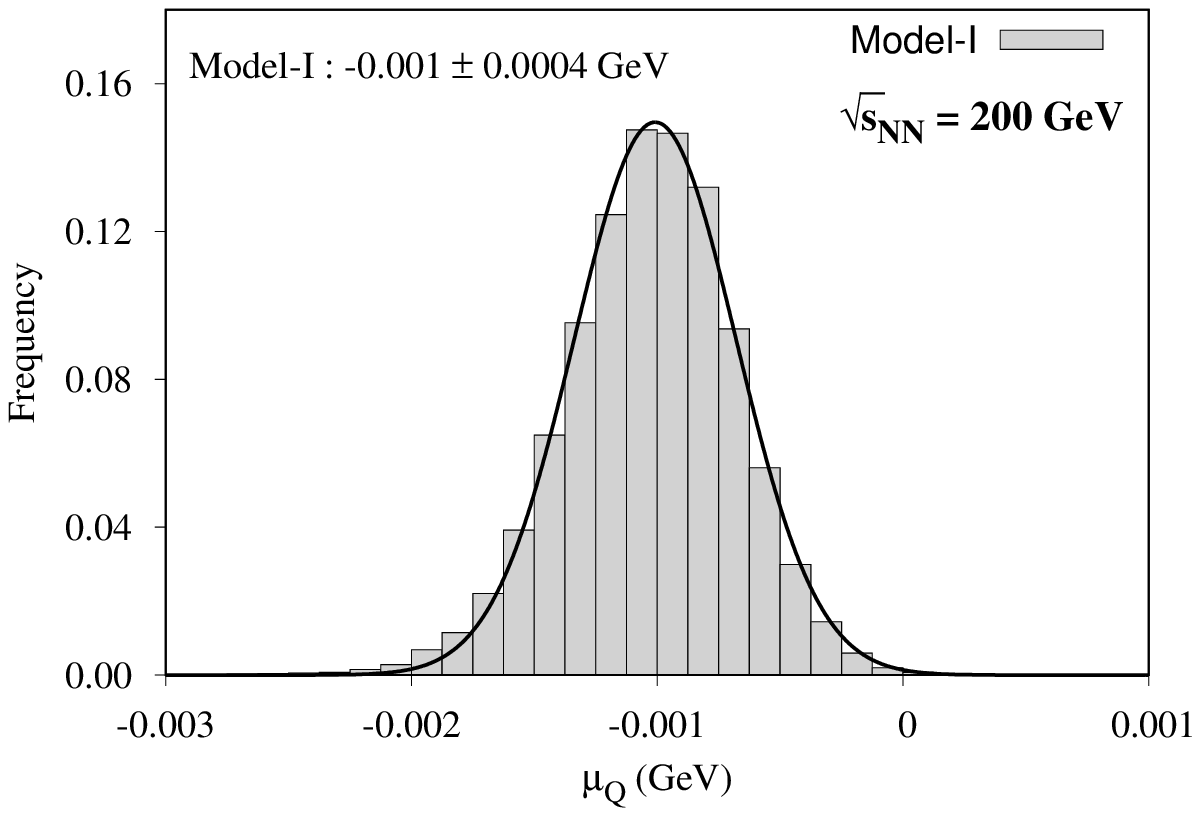}}
\label{fg.muqRHICwc}}
\subfloat[]{
{\includegraphics[scale=0.7]{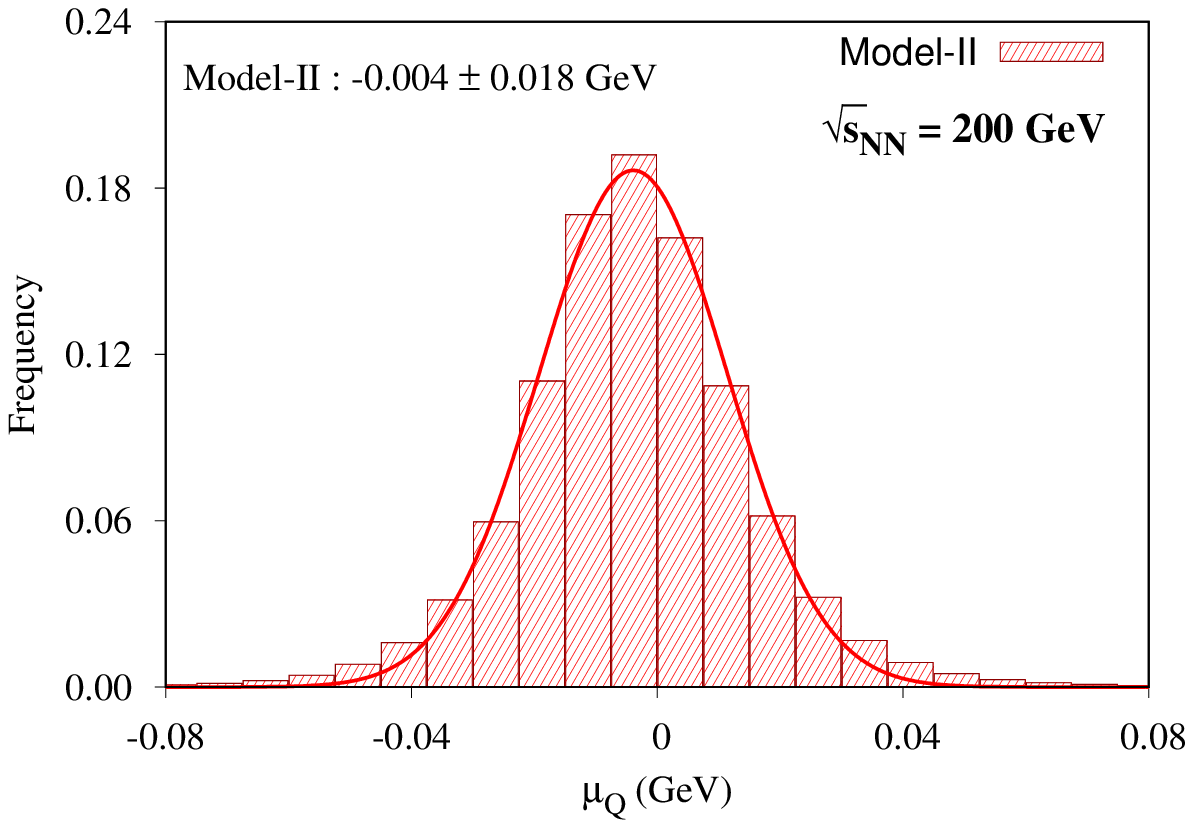}}
\label{fg.muqRHICwoc}}\\
\subfloat[]{
{\includegraphics[scale=0.7]{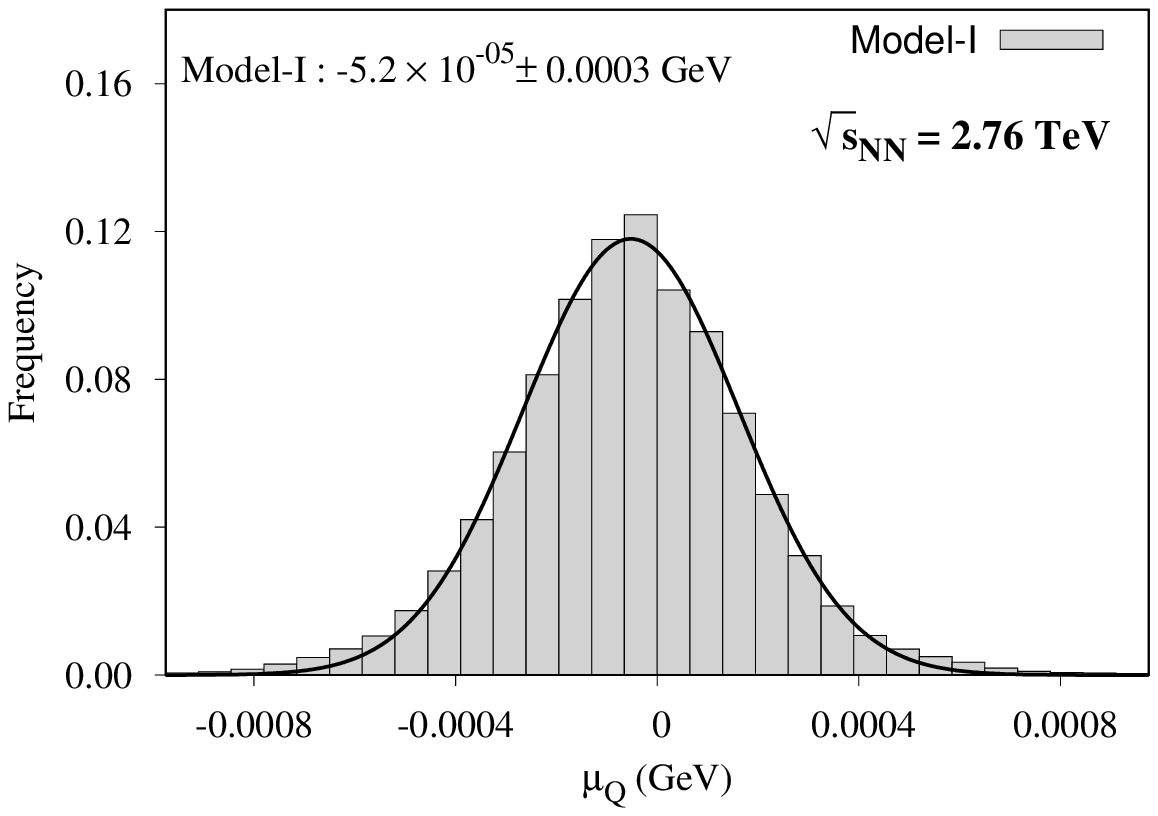}}
\label{fg.muqLHCwc}}
\subfloat[]{
{\includegraphics[scale=0.7]{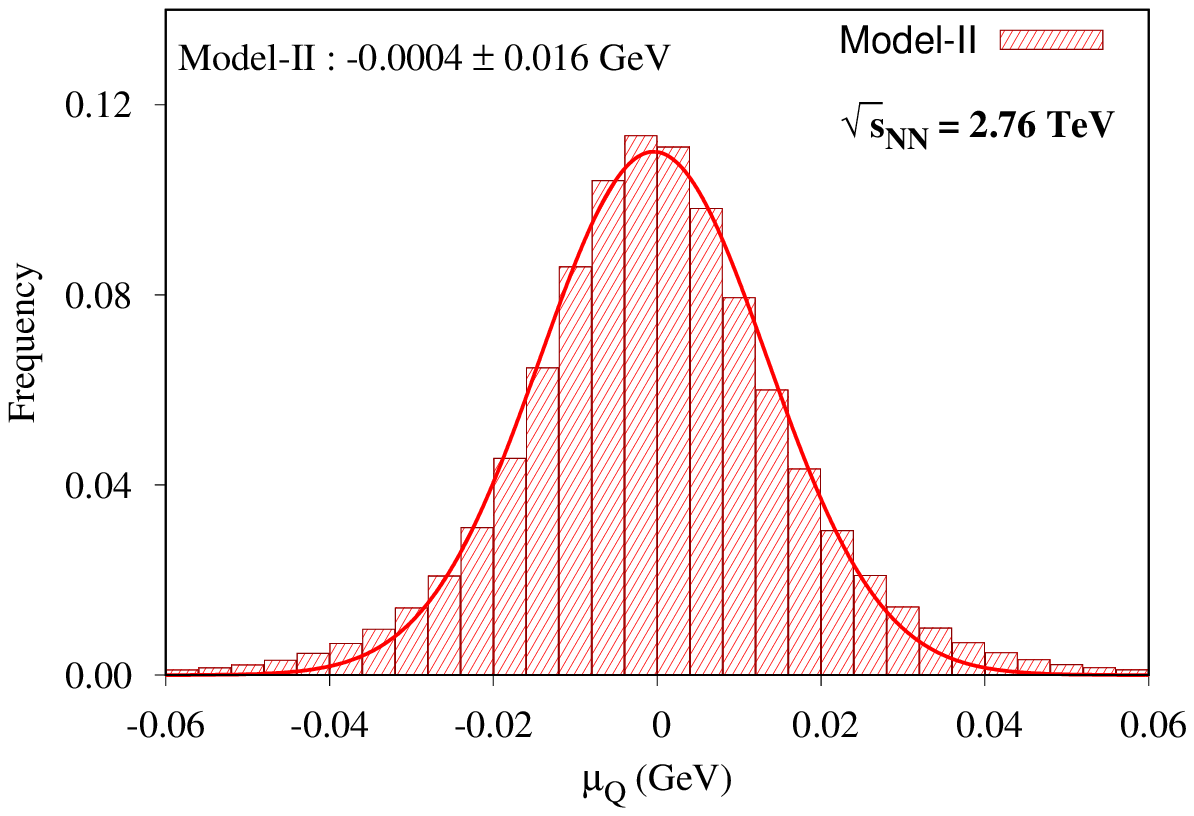}}
\label{fg.muqLHCwoc}}\\
\caption{
\label{fg.musmuqLHC}
The distribution of electric charge chemical potential at chemical freeze-out for collision energies $\sqrt{s_{NN}}=200$~GeV (RHIC) and $\sqrt{s_{NN}}=2.76$~TeV (LHC). Also shown are the fitted Gaussian curve and the values of extracted electric charge chemical potential at freeze-out with systematic error bars are listed, for both Model-I and Model-II.}
\end{figure*}
%%%%%%%%%%%%%%%%%%%%%%%%

Furthermore, as shown in Figs.~\ref{fg.muqRHICwc}~and~\ref{fg.muqRHICwoc}, the numerical value of $\mu_Q$ is small. Especially for LHC energy, as shown in Figs.~\ref{fg.muqLHCwc}~and~\ref{fg.muqLHCwoc}, it is vanishingly small. Therefore, exclusion of net baryon to net charge constraint in Model-II result in fluctuations. Only the $\chi^2$ analysis using these parameters can confirm the reliability of Model-II, which will be discussed later. 

%%%%%%%%%%%%%%%%%%%%%%%%
\begin{table*}[!htb]\centering
%\scalebox{0.9}
{\begin{tabular}{|c|c|c|c|c|cc|}
\hline
& \\[-2.6 ex]
Parameters & Model-I & Model-II & Andronic et. al~\cite{Andronic:2005yp} & STAR BES~\cite{Adamczyk:2017iwn} & \multicolumn{2}{c|}{Bhattacharyya et. al~\cite{Bhattacharyya:2019cer}}~\\[1 ex]
& \\[-4 ex]
for RHIC & ~ & ~ &~& ~ &WC~&~WOC~\\[1 ex]
\hline
\hline
& \\[-2 ex]
$T$ (GeV) & ~$0.169 \pm 0.003$~ & ~$0.169 \pm 0.004$~ & ~$0.160 \pm 0.006$~ & ~$0.164 \pm 0.005$~ & ~$0.165^{+0.008}_{-0.010}$~ & ~$0.166^{+0.010}_{-0.012}$~\\[1 ex]
\hline
\hline
& \\[-2 ex]
${\mu_B}$ (GeV) & ~$ 0.028 \pm 0.010$~ & ~$0.030 \pm 0.016$~ & ~$0.020 \pm 0.004$~ & ~$0.028 \pm 0.006$~ & ~$ 0.028^{+0.012}_{-0.015}$~ &~$ 0.032^{+0.010}_{-0.017}$~\\[1 ex]
\hline
\hline
& \\[-2 ex]
${\mu_S}$ (GeV) & ~$0.007 \pm 0.003 $~ & ~$0.009 \pm 0.010$~ &  & ~$0.006 \pm 0.004$~ & ~$0.007^{+0.002}_{-0.003}$~& ~$0.009^{+0.006}_{-0.008}$~\\[1 ex]
\hline
\hline
& \\[-2 ex]
${\mu_Q}$ (GeV) & ~$-0.001 \pm 0.0004$~ & ~$-0.004 \pm 0.018$~ & & &~$-0.0008^{+0.0004}_{-0.0003}$~&~$-0.005^{+0.008}_{-0.004} $~\\[1 ex]
\hline
\end{tabular}}
\caption{Comparison of the values of freeze-out parameters with recent literatures for $\sqrt{s_{NN}}=200$~GeV at RHIC.}
\label{tb.RHIC}
\end{table*}
%%%%%%%%%%%%%%%%%%%%%%%%

%%%%%%%%%%%%%%%%%%%%%%%%
\begin{table*}[!htb]\centering%\scalebox{0.9}
{\begin{tabular}{|c|c|c|c|cc|}
\hline
& \\[-2.6 ex]
Parameters & Model-I & Model-II & Andronic et. al~\cite{Andronic:2017pug} & \multicolumn{2}{c|}{Bhattacharyya et. al~\cite{Bhattacharyya:2019cer}}~\\[1 ex]
& \\[-4 ex]
for LHC & ~ & ~& ~ &WC~&~WOC~\\[1 ex]
\hline
\hline
& \\[-2 ex]
$T$ (GeV) & ~$0.155 \pm 0.003$~ & ~$0.155 \pm 0.003$~ & ~$0.157 \pm 0.002$~ & ~$0.152^{+0.008}_{-0.006}$~ & ~$0.152^{+0.007}_{-0.005}$~\\[1 ex]
\hline
\hline
& \\[-2 ex]
${\mu_B}$ (GeV) & ~$0.002 \pm 0.009$~ & ~$0.002 \pm 0.015$~ & ~$0.007 \pm 0.004$~& ~$ 0.003^{+0.013}_{-0.013}$~& ~$ 0.006^{+0.011}_{-0.016}$~\\[1 ex]
\hline
\hline
& \\[-2 ex]
${\mu_S}$ (GeV) & ~$0.0006 \pm 0.0018$~ & ~$0.0005 \pm 0.0089$~ & & ~$0.0005^{+0.0023}_{-0.0028}$~& ~$0.002^{+0.007}_{-0.012}$
~\\[1 ex]
\hline
\hline
& \\[-2 ex]
${\mu_Q}$ (GeV) & $-0.00005 \pm 0.0002$~ & ~$-0.0004 \pm 0.016 $~ & &${-0.00005}^{+0.0003}_{-0.0003}$~&~$-0.003 ^{+0.004}_{-0.002}$~\\[1 ex]
\hline
\end{tabular}}
\caption{Comparison of the values of freeze-out parameters with recent literatures for $\sqrt{s_{NN}}=2.76$~TeV at LHC.}
\label{tb.LHC}
\end{table*}
%%%%%%%%%%%%%%%%%%%%%%%%

%%%%%%%%%%%%%%%%%%%%%%%%
\begin{figure*}[!htb]
\centering
\subfloat[]{
{\includegraphics[scale=0.7]{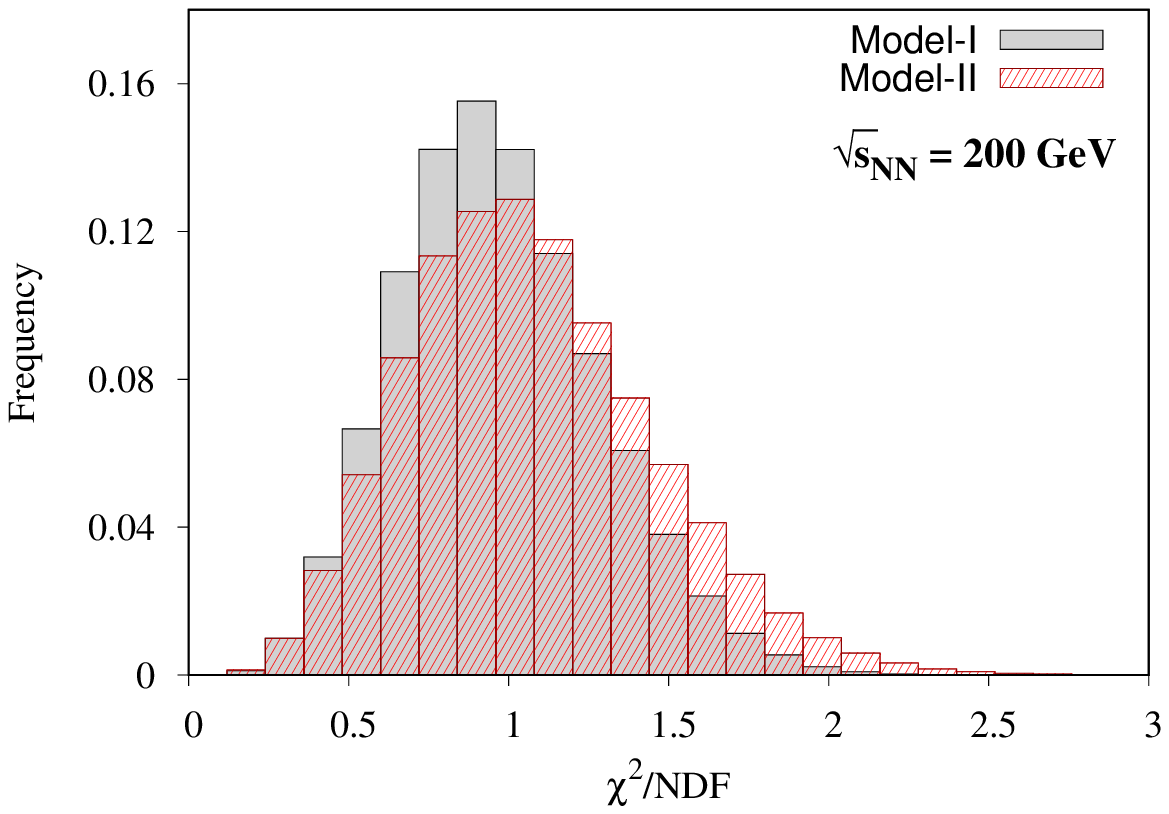}}
\label{fg.chiRHIC}}
\subfloat[]{
{\includegraphics[scale=0.7]{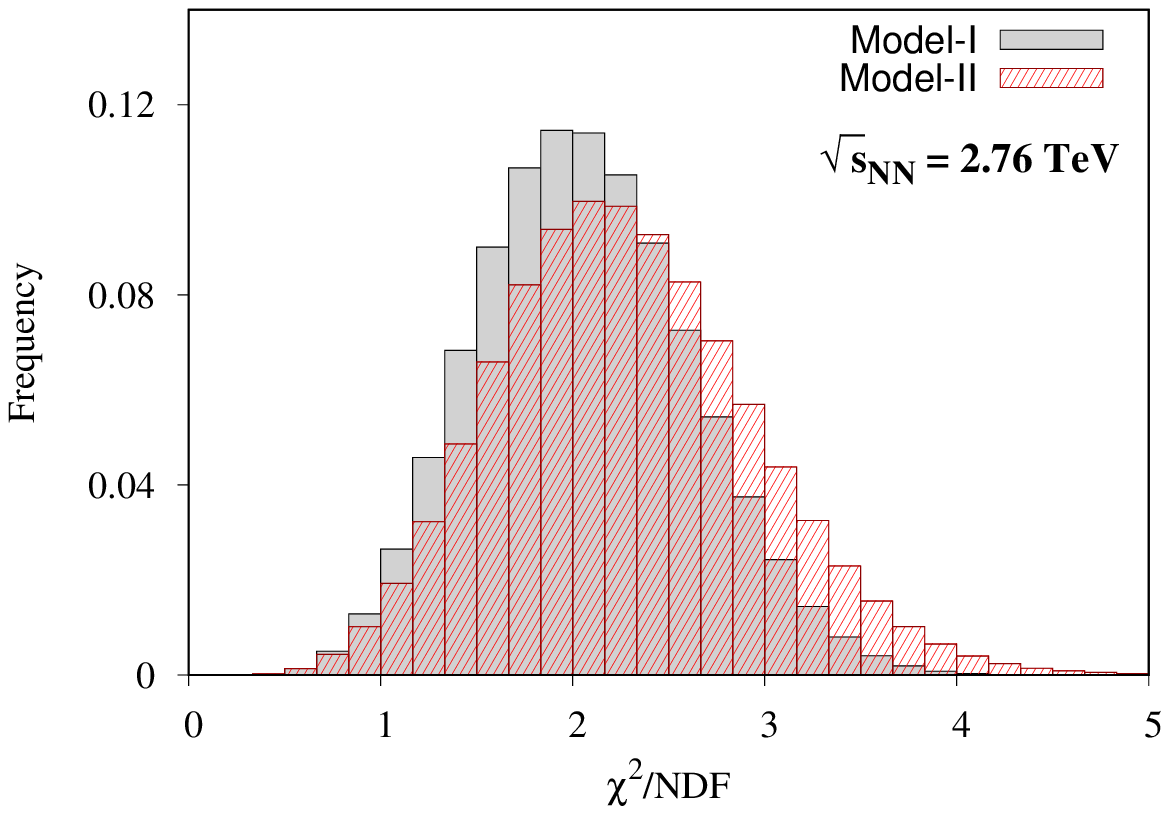}}
\label{fg.chiLHC}}\\
\caption{
The histogram distribution of $\chi^2$/NDF for all possible independent sets for collision energies $\sqrt{s_{NN}}=200$~GeV (RHIC) and $\sqrt{s_{NN}}=2.76$~TeV (LHC), for both Model-I and Model-II.}
\label{fg.chi}
\end{figure*}
%%%%%%%%%%%%%%%%%%%%%%%%

For comparison, the freeze-out parameters from relevant literature are also documented in Tables~\ref{tb.RHIC}~and~\ref{tb.LHC} for collision energies of $\sqrt{s_{NN}}=200$~GeV at RHIC and $\sqrt{s_{NN}}=2.76$~TeV at LHC, respectively. Specifically we have used results from Refs.~\cite{Andronic:2005yp, Adamczyk:2017iwn, Bhattacharyya:2019cer} for RHIC energy and Refs.~\cite{Andronic:2017pug, Bhattacharyya:2019cer} for LHC energy. For RHIC, we obtain a slightly larger value of the extracted freeze-out temperature. Nevertheless, we find that our extracted values of freeze-out parameters are more or less in agreement with those reported earlier in literature. However, one must keep in mind that there are significant differences in the analysis procedure and computation techniques employed here. The error bars considered in Ref.~\cite{Andronic:2005yp} are obtained by varying the $\chi^2$ by unity, which accounts for the fitting error. In Ref.~\cite{Adamczyk:2017iwn}, $\mu_Q$ was assumed to be vanishingly small and therefore neglected. However, an additional parameter $\gamma_s$ was introduced to account for the possible deviation of strange particle abundances from chemical equilibrium. The only set of ratios that are used in the analysis of Ref.~\cite{Adamczyk:2017iwn} are ${\pi}^{-}/{\pi}^{+}$, $k^{-}/k{+}$, ${\bar{p}}/p$, ${\bar{\Lambda}}/{\Lambda}$, ${\Xi}^{+}/{\Xi}^{-}$, $k^{-}/{\pi}^{-}$, ${\bar{p}}/{\pi}^{-}$, ${\Lambda}/{\pi}^{-}$ and ${\Xi}^{+}/{\pi}^{-}$. They have also studied different sources of systematic uncertainties, which propagates from yield to freeze-out parameters. In Ref.~\cite{Bhattacharyya:2019cer} the systematic error bars has been calculated with nine independent sets, considering the usual constraint of strong charges (WC) and without constrain relations (WOC) of Eq.~\eqref{eq:nbq} and Eq.~\eqref{eq:ns}. In Ref.~\cite{Andronic:2017pug}, the authors have considered system volume as a parameter and the standard deviations quoted there comes only from experimental uncertainties.

Reduced $\chi^2$ i.e. $\chi^2$/NDF gives a measure of goodness of fit, where NDF stands for the number of degrees of freedom. The usual practice is to choose a specific independent set so that the $\chi^2$ is equal to the number of degrees of freedom of the system. The histogram distribution of reduced $\chi^2$ corresponding to all possible independent sets for $\sqrt{s_{NN}}=2.76$~TeV (LHC) and $\sqrt{s_{NN}}=200$~GeV (RHIC) are shown for both Model-I and Model-II in Fig.~\ref{fg.chi}. We see that $\chi^2$/NDF for both the models differ only slightly. It is interesting to see that the distribution is peaked at $\chi^2/{\rm NDF}\simeq1$ for RHIC as shown in Fig.~\ref{fg.chiRHIC} indicating a good fit. On the other hand, from Fig.~\ref{fg.chiLHC}, we see that $\chi^2/{\rm NDF}\simeq2.2$ for LHC. This suggests that quality of fit at LHC may not be as good as that at RHIC. Anyhow, this result is not surprising as it has been pointed out in existing literature that reduced $\chi^2$ for LHC is quite large, if we consider the equilibration of all the hadrons simultaneously within the framework of the conserved charges~\cite{Stachel:2013zma,Chatterjee:2015fua}. However,  the usual practice is to minimize $\chi^2$ for a specific independent set and large value of $\chi^2$/NDF for that particular set does not necessarily be alarming. But, we have done the study for all possible independent sets and observe that the peak of the reduced $\chi^2$ distribution lies near 2.2, for both the models. Thus, the thermal model analysis for LHC data demands further study.

%%%%%%%%%%%%%%%%%%%%%%%%%%%%%%%%%%%%%%%%%%%%%%%%%
\section{Summary and outlook}
\label{sec:summary}

It has been conjectured in existing literature that $\chi^2$ fit of experimentally measured hadronic yields, or multiplicity ratios, within thermal model framework can successfully extract chemical freeze-out parameters to a certain extent. It is evident that, there are many possible sources of systematic uncertainties that can affect the analysis. The usual practice is to choose a specific set of independent ratios for the analysis, such that $\chi^2$/NDF is close to one. This is one of the sources of systematic uncertainty as the equilibration parameters are biased to the chosen particle ratios. Estimation of these uncertainties are important because it would shed light on the reliability of such thermal models where we have assumed a grand canonical ensemble with thermodynamic equilibration of all the hadrons emerging from the heavy-ion collision experiments.

In this article, we have provided an elegant method to quantify such systematic uncertainties in the chemical freeze-out parameters extracted from thermal model analysis of hadron multiplicity ratios. We have identified the enumeration problem of independent sets to a well known problem in combinatorial optimization and graph theory. By implementing the minimum spanning tree algorithm, we have generated all possible independent sets. This corresponds to the number of all different trees on $N$ vertices resulting in generation of $N^{N-2}$ independent sets as expected from Cayley’s formula.

We have performed the $\chi^2$ minimization on each sets to obtain freeze-out parameters corresponding to yield ratios of experimental data at collision energies of $\sqrt{s_{NN}}=200$~GeV at RHIC and $\sqrt{s_{NN}}=2.76$~TeV at LHC. Since the number of sets are extremely large ($10^8$), we obtain a distribution of these parameters. From these distributions, we estimate the mean value of the said parameter along with the quantitative measure of systematic uncertainty. Model-II seems to have larger systematic uncertainty than Model-I, which is expected because Model-II has one extra free parameter. But the $\chi^2$/NDF distribution for RHIC is peaked around one which ensures that the systematics are under control for both the models. One may be tempted to conclude that a  global equilibration scenario is achieved for RHIC where equilibration of all hadrons occurs simultaneously. On the other hand, for LHC, we found that although the parameters seem to be in good agreement with existing literature, the value where of $\chi^2$/NDF distribution is peaked is away from one. This leads to a serious challenge about the applicability of the thermal model (in the present form) at LHC energy.

At this juncture, we would like to emphasize that there are no empirical reasons to believe that the simplistic statistical hadronization models can explain the entire physical scenario near the cross-over region~\cite{Huovinen:2017ogf, Huovinen:2018ziu}. The uncertainties in the prediction at LHC energy, leads to study of alternative freeze-out schemes, such as~\cite{Steinheimer:2012rd, Petran:2013lja, Becattini:2014hla}. Moreover, even within thermal model analysis, there is a possibility of mass dependent or flavour dependent sequential freeze-out scenario. The intrinsic limitations of these statistical models are that one does not have the precise knowledge of how different ratios are dependent on the physical scenario. On the other hand, our method of estimating systematic uncertainties by generating all possible sets of independent ratios is quite general and is applicable also for data sets with bias. For instance, if one could estimate the bias on the ratios then the present method could be employed to fit all possible sets and calculate the weighted average of the fitted parameters and corresponding standard deviations. In absence of that knowledge, the present analysis is the best one can do to extract systematic errors which we have estimated in the present article.

Looking forward, the framework presented here can be applied to other collision energies at RHIC and LHC in order to discern whether the extraction of chemical freeze-out parameters are biased by the choice of yield ratios. Moreover, the current framework can be easily extended to physics analysis using double/multiple ratios which will have important implications for sequential freeze-out models. We leave these for future work.

%%%%%%%%%%%%%%%%%%%%%%%%%%%%%%%%%%%%%%%%%%%%%%%%%
\section{Acknowledgements}
S.B. thanks Deeptak Biswas and Rajarshi Ray for helpful discussions and acknowledges NISER for kind hospitality. A.J. thanks Tuhin Ghosh and Kush Saha for their help in using the VIRGO computing cluster at NISER. S.R. thanks Rishiraj Bhattacharyya and Himadri Nayak for fruitful discussions. S.B. was supported by CSIR and DST. A.J. and S.R. are supported in part by the DST-INSPIRE faculty award under Grant No. DST/INSPIRE/04/2017/000038 and Grant No. DST/INSPIRE/04/2016/000215, respectively. 
%%%%%%%%%%%%%%%%%%%%%%%%%%%%%%%%%%%%%%%%%%%%%%%%%

\bibliography{ref}

%merlin.mbs apsrev4-1.bst 2010-07-25 4.21a (PWD, AO, DPC) hacked
%Control: key (0)
%Control: author (8) initials jnrlst
%Control: editor formatted (1) identically to author
%Control: production of article title (-1) disabled
%Control: page (0) single
%Control: year (1) truncated
%Control: production of eprint (0) enabled
\begin{thebibliography}{94}%
\makeatletter
\providecommand \@ifxundefined [1]{%
 \@ifx{#1\undefined}
}%
\providecommand \@ifnum [1]{%
 \ifnum #1\expandafter \@firstoftwo
 \else \expandafter \@secondoftwo
 \fi
}%
\providecommand \@ifx [1]{%
 \ifx #1\expandafter \@firstoftwo
 \else \expandafter \@secondoftwo
 \fi
}%
\providecommand \natexlab [1]{#1}%
\providecommand \enquote  [1]{``#1''}%
\providecommand \bibnamefont  [1]{#1}%
\providecommand \bibfnamefont [1]{#1}%
\providecommand \citenamefont [1]{#1}%
\providecommand \href@noop [0]{\@secondoftwo}%
\providecommand \href [0]{\begingroup \@sanitize@url \@href}%
\providecommand \@href[1]{\@@startlink{#1}\@@href}%
\providecommand \@@href[1]{\endgroup#1\@@endlink}%
\providecommand \@sanitize@url [0]{\catcode `\\12\catcode `\$12\catcode
  `\&12\catcode `\#12\catcode `\^12\catcode `\_12\catcode `\%12\relax}%
\providecommand \@@startlink[1]{}%
\providecommand \@@endlink[0]{}%
\providecommand \url  [0]{\begingroup\@sanitize@url \@url }%
\providecommand \@url [1]{\endgroup\@href {#1}{\urlprefix }}%
\providecommand \urlprefix  [0]{URL }%
\providecommand \Eprint [0]{\href }%
\providecommand \doibase [0]{http://dx.doi.org/}%
\providecommand \selectlanguage [0]{\@gobble}%
\providecommand \bibinfo  [0]{\@secondoftwo}%
\providecommand \bibfield  [0]{\@secondoftwo}%
\providecommand \translation [1]{[#1]}%
\providecommand \BibitemOpen [0]{}%
\providecommand \bibitemStop [0]{}%
\providecommand \bibitemNoStop [0]{.\EOS\space}%
\providecommand \EOS [0]{\spacefactor3000\relax}%
\providecommand \BibitemShut  [1]{\csname bibitem#1\endcsname}%
\let\auto@bib@innerbib\@empty
%</preamble>
\bibitem [{\citenamefont {Fermi}(1950)}]{Fermi:1950jd}%
  \BibitemOpen
  \bibfield  {author} {\bibinfo {author} {\bibfnamefont {E.}~\bibnamefont
  {Fermi}},\ }\href {\doibase 10.1143/PTP.5.570} {\bibfield  {journal}
  {\bibinfo  {journal} {Prog. Theor. Phys.}\ }\textbf {\bibinfo {volume} {5}},\
  \bibinfo {pages} {570} (\bibinfo {year} {1950})}\BibitemShut {NoStop}%
%%CITATION = PTPKA,5,570;%%
\bibitem [{\citenamefont {Pomeranchuk}(1951)}]{Pomeranchuk:1951ey}%
  \BibitemOpen
  \bibfield  {author} {\bibinfo {author} {\bibfnamefont {I.~{\relax Ya}.}\
  \bibnamefont {Pomeranchuk}},\ }\href@noop {} {\bibfield  {journal} {\bibinfo
  {journal} {Dokl. Akad. Nauk Ser. Fiz.}\ }\textbf {\bibinfo {volume} {78}},\
  \bibinfo {pages} {889} (\bibinfo {year} {1951})},\ \bibinfo {note}
  {[,67(1951)]}\BibitemShut {NoStop}%
%%CITATION = DANKA,78,889;%%
\bibitem [{\citenamefont {Landau}(1953)}]{Landau:1953gs}%
  \BibitemOpen
  \bibfield  {author} {\bibinfo {author} {\bibfnamefont {L.~D.}\ \bibnamefont
  {Landau}},\ }\href@noop {} {\bibfield  {journal} {\bibinfo  {journal} {Izv.
  Akad. Nauk Ser. Fiz.}\ }\textbf {\bibinfo {volume} {17}},\ \bibinfo {pages}
  {51} (\bibinfo {year} {1953})}\BibitemShut {NoStop}%
%%CITATION = IANFA,17,51;%%
\bibitem [{\citenamefont {Belenkij}\ and\ \citenamefont
  {Landau}(1956)}]{Belenkij:1956cd}%
  \BibitemOpen
  \bibfield  {author} {\bibinfo {author} {\bibfnamefont {S.~Z.}\ \bibnamefont
  {Belenkij}}\ and\ \bibinfo {author} {\bibfnamefont {L.~D.}\ \bibnamefont
  {Landau}},\ }\href {\doibase 10.1007/BF02745507} {\bibfield  {journal}
  {\bibinfo  {journal} {Nuovo Cim. Suppl.}\ }\textbf {\bibinfo {volume}
  {3S10}},\ \bibinfo {pages} {15} (\bibinfo {year} {1956})},\ \bibinfo {note}
  {[Usp. Fiz. Nauk56,309(1955)]}\BibitemShut {NoStop}%
%%CITATION = NUCUA,3S10,15;%%
\bibitem [{\citenamefont {Hagedorn}\ and\ \citenamefont
  {Rafelski}(1980)}]{Hagedorn:1980kb}%
  \BibitemOpen
  \bibfield  {author} {\bibinfo {author} {\bibfnamefont {R.}~\bibnamefont
  {Hagedorn}}\ and\ \bibinfo {author} {\bibfnamefont {J.}~\bibnamefont
  {Rafelski}},\ }\href {\doibase 10.1016/0370-2693(80)90566-3} {\bibfield
  {journal} {\bibinfo  {journal} {Phys. Lett.}\ }\textbf {\bibinfo {volume}
  {97B}},\ \bibinfo {pages} {136} (\bibinfo {year} {1980})}\BibitemShut
  {NoStop}%
%%CITATION = PHLTA,97B,136;%%
\bibitem [{\citenamefont {Rischke}\ \emph {et~al.}(1991)\citenamefont
  {Rischke}, \citenamefont {Gorenstein}, \citenamefont {Stoecker},\ and\
  \citenamefont {Greiner}}]{Rischke:1991ke}%
  \BibitemOpen
  \bibfield  {author} {\bibinfo {author} {\bibfnamefont {D.~H.}\ \bibnamefont
  {Rischke}}, \bibinfo {author} {\bibfnamefont {M.~I.}\ \bibnamefont
  {Gorenstein}}, \bibinfo {author} {\bibfnamefont {H.}~\bibnamefont
  {Stoecker}}, \ and\ \bibinfo {author} {\bibfnamefont {W.}~\bibnamefont
  {Greiner}},\ }\href {\doibase 10.1007/BF01548574} {\bibfield  {journal}
  {\bibinfo  {journal} {Z. Phys.}\ }\textbf {\bibinfo {volume} {C51}},\
  \bibinfo {pages} {485} (\bibinfo {year} {1991})}\BibitemShut {NoStop}%
%%CITATION = ZEPYA,C51,485;%%
\bibitem [{\citenamefont {Cleymans}\ \emph {et~al.}(1993)\citenamefont
  {Cleymans}, \citenamefont {Gorenstein}, \citenamefont {Stalnacke},\ and\
  \citenamefont {Suhonen}}]{Cleymans:1992jz}%
  \BibitemOpen
  \bibfield  {author} {\bibinfo {author} {\bibfnamefont {J.}~\bibnamefont
  {Cleymans}}, \bibinfo {author} {\bibfnamefont {M.~I.}\ \bibnamefont
  {Gorenstein}}, \bibinfo {author} {\bibfnamefont {J.}~\bibnamefont
  {Stalnacke}}, \ and\ \bibinfo {author} {\bibfnamefont {E.}~\bibnamefont
  {Suhonen}},\ }\href {\doibase 10.1088/0031-8949/48/3/004} {\bibfield
  {journal} {\bibinfo  {journal} {Phys. Scripta}\ }\textbf {\bibinfo {volume}
  {48}},\ \bibinfo {pages} {277} (\bibinfo {year} {1993})}\BibitemShut
  {NoStop}%
%%CITATION = PHSTB,48,277;%%
\bibitem [{\citenamefont {Braun-Munzinger}\ \emph {et~al.}(1995)\citenamefont
  {Braun-Munzinger}, \citenamefont {Stachel}, \citenamefont {Wessels},\ and\
  \citenamefont {Xu}}]{BraunMunzinger:1994xr}%
  \BibitemOpen
  \bibfield  {author} {\bibinfo {author} {\bibfnamefont {P.}~\bibnamefont
  {Braun-Munzinger}}, \bibinfo {author} {\bibfnamefont {J.}~\bibnamefont
  {Stachel}}, \bibinfo {author} {\bibfnamefont {J.~P.}\ \bibnamefont
  {Wessels}}, \ and\ \bibinfo {author} {\bibfnamefont {N.}~\bibnamefont {Xu}},\
  }\href {\doibase 10.1016/0370-2693(94)01534-J} {\bibfield  {journal}
  {\bibinfo  {journal} {Phys. Lett.}\ }\textbf {\bibinfo {volume} {B344}},\
  \bibinfo {pages} {43} (\bibinfo {year} {1995})},\ \Eprint
  {http://arxiv.org/abs/nucl-th/9410026} {arXiv:nucl-th/9410026 [nucl-th]}
  \BibitemShut {NoStop}%
%%CITATION = NUCL-TH/9410026;%%
\bibitem [{\citenamefont {Cleymans}\ \emph {et~al.}(1997)\citenamefont
  {Cleymans}, \citenamefont {Elliott}, \citenamefont {Satz},\ and\
  \citenamefont {Thews}}]{Cleymans:1996cd}%
  \BibitemOpen
  \bibfield  {author} {\bibinfo {author} {\bibfnamefont {J.}~\bibnamefont
  {Cleymans}}, \bibinfo {author} {\bibfnamefont {D.}~\bibnamefont {Elliott}},
  \bibinfo {author} {\bibfnamefont {H.}~\bibnamefont {Satz}}, \ and\ \bibinfo
  {author} {\bibfnamefont {R.~L.}\ \bibnamefont {Thews}},\ }\href {\doibase
  10.1007/s002880050393} {\bibfield  {journal} {\bibinfo  {journal} {Z. Phys.}\
  }\textbf {\bibinfo {volume} {C74}},\ \bibinfo {pages} {319} (\bibinfo {year}
  {1997})},\ \Eprint {http://arxiv.org/abs/nucl-th/9603004}
  {arXiv:nucl-th/9603004 [nucl-th]} \BibitemShut {NoStop}%
%%CITATION = NUCL-TH/9603004;%%
\bibitem [{\citenamefont {Yen}\ \emph {et~al.}(1997)\citenamefont {Yen},
  \citenamefont {Gorenstein}, \citenamefont {Greiner},\ and\ \citenamefont
  {Yang}}]{Yen:1997rv}%
  \BibitemOpen
  \bibfield  {author} {\bibinfo {author} {\bibfnamefont {G.~D.}\ \bibnamefont
  {Yen}}, \bibinfo {author} {\bibfnamefont {M.~I.}\ \bibnamefont {Gorenstein}},
  \bibinfo {author} {\bibfnamefont {W.}~\bibnamefont {Greiner}}, \ and\
  \bibinfo {author} {\bibfnamefont {S.-N.}\ \bibnamefont {Yang}},\ }\href
  {\doibase 10.1103/PhysRevC.56.2210} {\bibfield  {journal} {\bibinfo
  {journal} {Phys. Rev.}\ }\textbf {\bibinfo {volume} {C56}},\ \bibinfo {pages}
  {2210} (\bibinfo {year} {1997})},\ \Eprint
  {http://arxiv.org/abs/nucl-th/9711062} {arXiv:nucl-th/9711062 [nucl-th]}
  \BibitemShut {NoStop}%
%%CITATION = NUCL-TH/9711062;%%
\bibitem [{\citenamefont {Heinz}(1999)}]{Heinz:1998st}%
  \BibitemOpen
  \bibfield  {author} {\bibinfo {author} {\bibfnamefont {U.~W.}\ \bibnamefont
  {Heinz}},\ }\href {\doibase 10.1088/0954-3899/25/2/014} {\bibfield  {journal}
  {\bibinfo  {journal} {J. Phys.}\ }\textbf {\bibinfo {volume} {G25}},\
  \bibinfo {pages} {263} (\bibinfo {year} {1999})},\ \Eprint
  {http://arxiv.org/abs/nucl-th/9810056} {arXiv:nucl-th/9810056 [nucl-th]}
  \BibitemShut {NoStop}%
%%CITATION = NUCL-TH/9810056;%%
\bibitem [{\citenamefont {Cleymans}\ and\ \citenamefont
  {Redlich}(1998)}]{Cleymans:1998fq}%
  \BibitemOpen
  \bibfield  {author} {\bibinfo {author} {\bibfnamefont {J.}~\bibnamefont
  {Cleymans}}\ and\ \bibinfo {author} {\bibfnamefont {K.}~\bibnamefont
  {Redlich}},\ }\href {\doibase 10.1103/PhysRevLett.81.5284} {\bibfield
  {journal} {\bibinfo  {journal} {Phys. Rev. Lett.}\ }\textbf {\bibinfo
  {volume} {81}},\ \bibinfo {pages} {5284} (\bibinfo {year} {1998})},\ \Eprint
  {http://arxiv.org/abs/nucl-th/9808030} {arXiv:nucl-th/9808030} \BibitemShut
  {NoStop}%
\bibitem [{\citenamefont {Braun-Munzinger}\ \emph {et~al.}(1999)\citenamefont
  {Braun-Munzinger}, \citenamefont {Heppe},\ and\ \citenamefont
  {Stachel}}]{BraunMunzinger:1999qy}%
  \BibitemOpen
  \bibfield  {author} {\bibinfo {author} {\bibfnamefont {P.}~\bibnamefont
  {Braun-Munzinger}}, \bibinfo {author} {\bibfnamefont {I.}~\bibnamefont
  {Heppe}}, \ and\ \bibinfo {author} {\bibfnamefont {J.}~\bibnamefont
  {Stachel}},\ }\href {\doibase 10.1016/S0370-2693(99)01076-X} {\bibfield
  {journal} {\bibinfo  {journal} {Phys. Lett.}\ }\textbf {\bibinfo {volume}
  {B465}},\ \bibinfo {pages} {15} (\bibinfo {year} {1999})},\ \Eprint
  {http://arxiv.org/abs/nucl-th/9903010} {arXiv:nucl-th/9903010 [nucl-th]}
  \BibitemShut {NoStop}%
%%CITATION = NUCL-TH/9903010;%%
\bibitem [{\citenamefont {Cleymans}\ and\ \citenamefont
  {Redlich}(1999)}]{Cleymans:1999st}%
  \BibitemOpen
  \bibfield  {author} {\bibinfo {author} {\bibfnamefont {J.}~\bibnamefont
  {Cleymans}}\ and\ \bibinfo {author} {\bibfnamefont {K.}~\bibnamefont
  {Redlich}},\ }\href {\doibase 10.1103/PhysRevC.60.054908} {\bibfield
  {journal} {\bibinfo  {journal} {Phys. Rev.}\ }\textbf {\bibinfo {volume}
  {C60}},\ \bibinfo {pages} {054908} (\bibinfo {year} {1999})},\ \Eprint
  {http://arxiv.org/abs/nucl-th/9903063} {arXiv:nucl-th/9903063 [nucl-th]}
  \BibitemShut {NoStop}%
%%CITATION = NUCL-TH/9903063;%%
\bibitem [{\citenamefont {Braun-Munzinger}\ \emph {et~al.}(2001)\citenamefont
  {Braun-Munzinger}, \citenamefont {Magestro}, \citenamefont {Redlich},\ and\
  \citenamefont {Stachel}}]{BraunMunzinger:2001ip}%
  \BibitemOpen
  \bibfield  {author} {\bibinfo {author} {\bibfnamefont {P.}~\bibnamefont
  {Braun-Munzinger}}, \bibinfo {author} {\bibfnamefont {D.}~\bibnamefont
  {Magestro}}, \bibinfo {author} {\bibfnamefont {K.}~\bibnamefont {Redlich}}, \
  and\ \bibinfo {author} {\bibfnamefont {J.}~\bibnamefont {Stachel}},\ }\href
  {\doibase 10.1016/S0370-2693(01)01069-3} {\bibfield  {journal} {\bibinfo
  {journal} {Phys. Lett.}\ }\textbf {\bibinfo {volume} {B518}},\ \bibinfo
  {pages} {41} (\bibinfo {year} {2001})},\ \Eprint
  {http://arxiv.org/abs/hep-ph/0105229} {arXiv:hep-ph/0105229 [hep-ph]}
  \BibitemShut {NoStop}%
%%CITATION = HEP-PH/0105229;%%
\bibitem [{\citenamefont {Becattini}\ \emph {et~al.}(2004)\citenamefont
  {Becattini}, \citenamefont {Gazdzicki}, \citenamefont {Keranen},
  \citenamefont {Manninen},\ and\ \citenamefont {Stock}}]{Becattini:2003wp}%
  \BibitemOpen
  \bibfield  {author} {\bibinfo {author} {\bibfnamefont {F.}~\bibnamefont
  {Becattini}}, \bibinfo {author} {\bibfnamefont {M.}~\bibnamefont
  {Gazdzicki}}, \bibinfo {author} {\bibfnamefont {A.}~\bibnamefont {Keranen}},
  \bibinfo {author} {\bibfnamefont {J.}~\bibnamefont {Manninen}}, \ and\
  \bibinfo {author} {\bibfnamefont {R.}~\bibnamefont {Stock}},\ }\href
  {\doibase 10.1103/PhysRevC.69.024905} {\bibfield  {journal} {\bibinfo
  {journal} {Phys. Rev.}\ }\textbf {\bibinfo {volume} {C69}},\ \bibinfo {pages}
  {024905} (\bibinfo {year} {2004})},\ \Eprint
  {http://arxiv.org/abs/hep-ph/0310049} {arXiv:hep-ph/0310049 [hep-ph]}
  \BibitemShut {NoStop}%
%%CITATION = HEP-PH/0310049;%%
\bibitem [{\citenamefont {Braun-Munzinger}\ \emph {et~al.}(2003)\citenamefont
  {Braun-Munzinger}, \citenamefont {Redlich},\ and\ \citenamefont
  {Stachel}}]{BraunMunzinger:2003zd}%
  \BibitemOpen
  \bibfield  {author} {\bibinfo {author} {\bibfnamefont {P.}~\bibnamefont
  {Braun-Munzinger}}, \bibinfo {author} {\bibfnamefont {K.}~\bibnamefont
  {Redlich}}, \ and\ \bibinfo {author} {\bibfnamefont {J.}~\bibnamefont
  {Stachel}},\ }\href@noop {} {\  (\bibinfo {year} {2003})},\ \Eprint
  {http://arxiv.org/abs/nucl-th/0304013} {arXiv:nucl-th/0304013 [nucl-th]}
  \BibitemShut {NoStop}%
%%CITATION = NUCL-TH/0304013;%%
\bibitem [{\citenamefont {Karsch}\ \emph {et~al.}(2003)\citenamefont {Karsch},
  \citenamefont {Redlich},\ and\ \citenamefont {Tawfik}}]{Karsch:2003zq}%
  \BibitemOpen
  \bibfield  {author} {\bibinfo {author} {\bibfnamefont {F.}~\bibnamefont
  {Karsch}}, \bibinfo {author} {\bibfnamefont {K.}~\bibnamefont {Redlich}}, \
  and\ \bibinfo {author} {\bibfnamefont {A.}~\bibnamefont {Tawfik}},\ }\href
  {\doibase 10.1016/j.physletb.2003.08.001} {\bibfield  {journal} {\bibinfo
  {journal} {Phys. Lett.}\ }\textbf {\bibinfo {volume} {B571}},\ \bibinfo
  {pages} {67} (\bibinfo {year} {2003})},\ \Eprint
  {http://arxiv.org/abs/hep-ph/0306208} {arXiv:hep-ph/0306208 [hep-ph]}
  \BibitemShut {NoStop}%
%%CITATION = HEP-PH/0306208;%%
\bibitem [{\citenamefont {Tawfik}(2005)}]{Tawfik:2004sw}%
  \BibitemOpen
  \bibfield  {author} {\bibinfo {author} {\bibfnamefont {A.}~\bibnamefont
  {Tawfik}},\ }\href {\doibase 10.1103/PhysRevD.71.054502} {\bibfield
  {journal} {\bibinfo  {journal} {Phys. Rev.}\ }\textbf {\bibinfo {volume}
  {D71}},\ \bibinfo {pages} {054502} (\bibinfo {year} {2005})},\ \Eprint
  {http://arxiv.org/abs/hep-ph/0412336} {arXiv:hep-ph/0412336 [hep-ph]}
  \BibitemShut {NoStop}%
%%CITATION = HEP-PH/0412336;%%
\bibitem [{\citenamefont {Becattini}\ \emph {et~al.}(2006)\citenamefont
  {Becattini}, \citenamefont {Manninen},\ and\ \citenamefont
  {Gazdzicki}}]{Becattini:2005xt}%
  \BibitemOpen
  \bibfield  {author} {\bibinfo {author} {\bibfnamefont {F.}~\bibnamefont
  {Becattini}}, \bibinfo {author} {\bibfnamefont {J.}~\bibnamefont {Manninen}},
  \ and\ \bibinfo {author} {\bibfnamefont {M.}~\bibnamefont {Gazdzicki}},\
  }\href {\doibase 10.1103/PhysRevC.73.044905} {\bibfield  {journal} {\bibinfo
  {journal} {Phys. Rev.}\ }\textbf {\bibinfo {volume} {C73}},\ \bibinfo {pages}
  {044905} (\bibinfo {year} {2006})},\ \Eprint
  {http://arxiv.org/abs/hep-ph/0511092} {arXiv:hep-ph/0511092 [hep-ph]}
  \BibitemShut {NoStop}%
%%CITATION = HEP-PH/0511092;%%
\bibitem [{\citenamefont {Andronic}\ \emph {et~al.}(2006)\citenamefont
  {Andronic}, \citenamefont {Braun-Munzinger},\ and\ \citenamefont
  {Stachel}}]{Andronic:2005yp}%
  \BibitemOpen
  \bibfield  {author} {\bibinfo {author} {\bibfnamefont {A.}~\bibnamefont
  {Andronic}}, \bibinfo {author} {\bibfnamefont {P.}~\bibnamefont
  {Braun-Munzinger}}, \ and\ \bibinfo {author} {\bibfnamefont {J.}~\bibnamefont
  {Stachel}},\ }\href {\doibase 10.1016/j.nuclphysa.2006.03.012} {\bibfield
  {journal} {\bibinfo  {journal} {Nucl. Phys.}\ }\textbf {\bibinfo {volume}
  {A772}},\ \bibinfo {pages} {167} (\bibinfo {year} {2006})},\ \Eprint
  {http://arxiv.org/abs/nucl-th/0511071} {arXiv:nucl-th/0511071 [nucl-th]}
  \BibitemShut {NoStop}%
%%CITATION = NUCL-TH/0511071;%%
\bibitem [{\citenamefont {Andronic}\ \emph {et~al.}(2009)\citenamefont
  {Andronic}, \citenamefont {Braun-Munzinger},\ and\ \citenamefont
  {Stachel}}]{Andronic:2008gu}%
  \BibitemOpen
  \bibfield  {author} {\bibinfo {author} {\bibfnamefont {A.}~\bibnamefont
  {Andronic}}, \bibinfo {author} {\bibfnamefont {P.}~\bibnamefont
  {Braun-Munzinger}}, \ and\ \bibinfo {author} {\bibfnamefont {J.}~\bibnamefont
  {Stachel}},\ }\href {\doibase 10.1016/j.physletb.2009.02.014,
  10.1016/j.physletb.2009.06.021} {\bibfield  {journal} {\bibinfo  {journal}
  {Phys. Lett.}\ }\textbf {\bibinfo {volume} {B673}},\ \bibinfo {pages} {142}
  (\bibinfo {year} {2009})},\ \bibinfo {note} {[Erratum: Phys.
  Lett.B678,516(2009)]},\ \Eprint {http://arxiv.org/abs/0812.1186}
  {arXiv:0812.1186 [nucl-th]} \BibitemShut {NoStop}%
%%CITATION = ARXIV:0812.1186;%%
\bibitem [{\citenamefont {Manninen}\ and\ \citenamefont
  {Becattini}(2008)}]{Manninen:2008mg}%
  \BibitemOpen
  \bibfield  {author} {\bibinfo {author} {\bibfnamefont {J.}~\bibnamefont
  {Manninen}}\ and\ \bibinfo {author} {\bibfnamefont {F.}~\bibnamefont
  {Becattini}},\ }\href {\doibase 10.1103/PhysRevC.78.054901} {\bibfield
  {journal} {\bibinfo  {journal} {Phys. Rev.}\ }\textbf {\bibinfo {volume}
  {C78}},\ \bibinfo {pages} {054901} (\bibinfo {year} {2008})},\ \Eprint
  {http://arxiv.org/abs/0806.4100} {arXiv:0806.4100 [nucl-th]} \BibitemShut
  {NoStop}%
%%CITATION = ARXIV:0806.4100;%%
\bibitem [{\citenamefont {Andronic}\ \emph {et~al.}(2012)\citenamefont
  {Andronic}, \citenamefont {Braun-Munzinger}, \citenamefont {Stachel},\ and\
  \citenamefont {Winn}}]{Andronic:2012ut}%
  \BibitemOpen
  \bibfield  {author} {\bibinfo {author} {\bibfnamefont {A.}~\bibnamefont
  {Andronic}}, \bibinfo {author} {\bibfnamefont {P.}~\bibnamefont
  {Braun-Munzinger}}, \bibinfo {author} {\bibfnamefont {J.}~\bibnamefont
  {Stachel}}, \ and\ \bibinfo {author} {\bibfnamefont {M.}~\bibnamefont
  {Winn}},\ }\href {\doibase 10.1016/j.physletb.2012.10.001} {\bibfield
  {journal} {\bibinfo  {journal} {Phys. Lett.}\ }\textbf {\bibinfo {volume}
  {B718}},\ \bibinfo {pages} {80} (\bibinfo {year} {2012})},\ \Eprint
  {http://arxiv.org/abs/1201.0693} {arXiv:1201.0693 [nucl-th]} \BibitemShut
  {NoStop}%
%%CITATION = ARXIV:1201.0693;%%
\bibitem [{\citenamefont {Tiwari}\ \emph {et~al.}(2012)\citenamefont {Tiwari},
  \citenamefont {Srivastava},\ and\ \citenamefont {Singh}}]{Tiwari:2011km}%
  \BibitemOpen
  \bibfield  {author} {\bibinfo {author} {\bibfnamefont {S.~K.}\ \bibnamefont
  {Tiwari}}, \bibinfo {author} {\bibfnamefont {P.~K.}\ \bibnamefont
  {Srivastava}}, \ and\ \bibinfo {author} {\bibfnamefont {C.~P.}\ \bibnamefont
  {Singh}},\ }\href {\doibase 10.1103/PhysRevC.85.014908} {\bibfield  {journal}
  {\bibinfo  {journal} {Phys. Rev.}\ }\textbf {\bibinfo {volume} {C85}},\
  \bibinfo {pages} {014908} (\bibinfo {year} {2012})},\ \Eprint
  {http://arxiv.org/abs/1111.2406} {arXiv:1111.2406 [hep-ph]} \BibitemShut
  {NoStop}%
%%CITATION = ARXIV:1111.2406;%%
\bibitem [{\citenamefont {Begun}\ \emph {et~al.}(2013)\citenamefont {Begun},
  \citenamefont {Gazdzicki},\ and\ \citenamefont {Gorenstein}}]{Begun:2012rf}%
  \BibitemOpen
  \bibfield  {author} {\bibinfo {author} {\bibfnamefont {V.~V.}\ \bibnamefont
  {Begun}}, \bibinfo {author} {\bibfnamefont {M.}~\bibnamefont {Gazdzicki}}, \
  and\ \bibinfo {author} {\bibfnamefont {M.~I.}\ \bibnamefont {Gorenstein}},\
  }\href {\doibase 10.1103/PhysRevC.88.024902} {\bibfield  {journal} {\bibinfo
  {journal} {Phys. Rev.}\ }\textbf {\bibinfo {volume} {C88}},\ \bibinfo {pages}
  {024902} (\bibinfo {year} {2013})},\ \Eprint {http://arxiv.org/abs/1208.4107}
  {arXiv:1208.4107 [nucl-th]} \BibitemShut {NoStop}%
%%CITATION = ARXIV:1208.4107;%%
\bibitem [{\citenamefont {Fu}(2013)}]{Fu:2013gga}%
  \BibitemOpen
  \bibfield  {author} {\bibinfo {author} {\bibfnamefont {J.}~\bibnamefont
  {Fu}},\ }\href {\doibase 10.1016/j.physletb.2013.04.018} {\bibfield
  {journal} {\bibinfo  {journal} {Phys. Lett.}\ }\textbf {\bibinfo {volume}
  {B722}},\ \bibinfo {pages} {144} (\bibinfo {year} {2013})}\BibitemShut
  {NoStop}%
%%CITATION = PHLTA,B722,144;%%
\bibitem [{\citenamefont {Tawfik}(2013)}]{Tawfik:2013eua}%
  \BibitemOpen
  \bibfield  {author} {\bibinfo {author} {\bibfnamefont {A.}~\bibnamefont
  {Tawfik}},\ }\href {\doibase 10.1103/PhysRevC.88.035203} {\bibfield
  {journal} {\bibinfo  {journal} {Phys. Rev.}\ }\textbf {\bibinfo {volume}
  {C88}},\ \bibinfo {pages} {035203} (\bibinfo {year} {2013})},\ \Eprint
  {http://arxiv.org/abs/1308.1712} {arXiv:1308.1712 [hep-ph]} \BibitemShut
  {NoStop}%
%%CITATION = ARXIV:1308.1712;%%
\bibitem [{\citenamefont {Garg}\ \emph {et~al.}(2013)\citenamefont {Garg},
  \citenamefont {Mishra}, \citenamefont {Netrakanti}, \citenamefont {Mohanty},
  \citenamefont {Mohanty}, \citenamefont {Singh},\ and\ \citenamefont
  {Xu}}]{Garg:2013ata}%
  \BibitemOpen
  \bibfield  {author} {\bibinfo {author} {\bibfnamefont {P.}~\bibnamefont
  {Garg}}, \bibinfo {author} {\bibfnamefont {D.~K.}\ \bibnamefont {Mishra}},
  \bibinfo {author} {\bibfnamefont {P.~K.}\ \bibnamefont {Netrakanti}},
  \bibinfo {author} {\bibfnamefont {B.}~\bibnamefont {Mohanty}}, \bibinfo
  {author} {\bibfnamefont {A.~K.}\ \bibnamefont {Mohanty}}, \bibinfo {author}
  {\bibfnamefont {B.~K.}\ \bibnamefont {Singh}}, \ and\ \bibinfo {author}
  {\bibfnamefont {N.}~\bibnamefont {Xu}},\ }\href {\doibase
  10.1016/j.physletb.2013.09.019} {\bibfield  {journal} {\bibinfo  {journal}
  {Phys. Lett.}\ }\textbf {\bibinfo {volume} {B726}},\ \bibinfo {pages} {691}
  (\bibinfo {year} {2013})},\ \Eprint {http://arxiv.org/abs/1304.7133}
  {arXiv:1304.7133 [nucl-ex]} \BibitemShut {NoStop}%
%%CITATION = ARXIV:1304.7133;%%
\bibitem [{\citenamefont {Bhattacharyya}\ \emph {et~al.}(2014)\citenamefont
  {Bhattacharyya}, \citenamefont {Das}, \citenamefont {Ghosh}, \citenamefont
  {Ray},\ and\ \citenamefont {Samanta}}]{Bhattacharyya:2013oya}%
  \BibitemOpen
  \bibfield  {author} {\bibinfo {author} {\bibfnamefont {A.}~\bibnamefont
  {Bhattacharyya}}, \bibinfo {author} {\bibfnamefont {S.}~\bibnamefont {Das}},
  \bibinfo {author} {\bibfnamefont {S.~K.}\ \bibnamefont {Ghosh}}, \bibinfo
  {author} {\bibfnamefont {R.}~\bibnamefont {Ray}}, \ and\ \bibinfo {author}
  {\bibfnamefont {S.}~\bibnamefont {Samanta}},\ }\href {\doibase
  10.1103/PhysRevC.90.034909} {\bibfield  {journal} {\bibinfo  {journal} {Phys.
  Rev.}\ }\textbf {\bibinfo {volume} {C90}},\ \bibinfo {pages} {034909}
  (\bibinfo {year} {2014})},\ \Eprint {http://arxiv.org/abs/1310.2793}
  {arXiv:1310.2793 [hep-ph]} \BibitemShut {NoStop}%
%%CITATION = ARXIV:1310.2793;%%
\bibitem [{\citenamefont {Albright}\ \emph {et~al.}(2014)\citenamefont
  {Albright}, \citenamefont {Kapusta},\ and\ \citenamefont
  {Young}}]{Albright:2014gva}%
  \BibitemOpen
  \bibfield  {author} {\bibinfo {author} {\bibfnamefont {M.}~\bibnamefont
  {Albright}}, \bibinfo {author} {\bibfnamefont {J.}~\bibnamefont {Kapusta}}, \
  and\ \bibinfo {author} {\bibfnamefont {C.}~\bibnamefont {Young}},\ }\href
  {\doibase 10.1103/PhysRevC.90.024915} {\bibfield  {journal} {\bibinfo
  {journal} {Phys. Rev.}\ }\textbf {\bibinfo {volume} {C90}},\ \bibinfo {pages}
  {024915} (\bibinfo {year} {2014})},\ \Eprint {http://arxiv.org/abs/1404.7540}
  {arXiv:1404.7540 [nucl-th]} \BibitemShut {NoStop}%
%%CITATION = ARXIV:1404.7540;%%
\bibitem [{\citenamefont {Kadam}\ and\ \citenamefont
  {Mishra}(2015)}]{Kadam:2015xsa}%
  \BibitemOpen
  \bibfield  {author} {\bibinfo {author} {\bibfnamefont {G.~P.}\ \bibnamefont
  {Kadam}}\ and\ \bibinfo {author} {\bibfnamefont {H.}~\bibnamefont {Mishra}},\
  }\href {\doibase 10.1103/PhysRevC.92.035203} {\bibfield  {journal} {\bibinfo
  {journal} {Phys. Rev.}\ }\textbf {\bibinfo {volume} {C92}},\ \bibinfo {pages}
  {035203} (\bibinfo {year} {2015})},\ \Eprint
  {http://arxiv.org/abs/1506.04613} {arXiv:1506.04613 [hep-ph]} \BibitemShut
  {NoStop}%
%%CITATION = ARXIV:1506.04613;%%
\bibitem [{\citenamefont {Kadam}\ and\ \citenamefont
  {Mishra}(2016)}]{Kadam:2015fza}%
  \BibitemOpen
  \bibfield  {author} {\bibinfo {author} {\bibfnamefont {G.~P.}\ \bibnamefont
  {Kadam}}\ and\ \bibinfo {author} {\bibfnamefont {H.}~\bibnamefont {Mishra}},\
  }\href {\doibase 10.1103/PhysRevC.93.025205} {\bibfield  {journal} {\bibinfo
  {journal} {Phys. Rev.}\ }\textbf {\bibinfo {volume} {C93}},\ \bibinfo {pages}
  {025205} (\bibinfo {year} {2016})},\ \Eprint
  {http://arxiv.org/abs/1509.06998} {arXiv:1509.06998 [hep-ph]} \BibitemShut
  {NoStop}%
%%CITATION = ARXIV:1509.06998;%%
\bibitem [{\citenamefont {Kadam}(2015)}]{Kadam:2015dda}%
  \BibitemOpen
  \bibfield  {author} {\bibinfo {author} {\bibfnamefont {G.~P.}\ \bibnamefont
  {Kadam}},\ }\href@noop {} {\  (\bibinfo {year} {2015})},\ \Eprint
  {http://arxiv.org/abs/1510.04371} {arXiv:1510.04371 [hep-ph]} \BibitemShut
  {NoStop}%
%%CITATION = ARXIV:1510.04371;%%
\bibitem [{\citenamefont {Albright}\ \emph {et~al.}(2015)\citenamefont
  {Albright}, \citenamefont {Kapusta},\ and\ \citenamefont
  {Young}}]{Albright:2015uua}%
  \BibitemOpen
  \bibfield  {author} {\bibinfo {author} {\bibfnamefont {M.}~\bibnamefont
  {Albright}}, \bibinfo {author} {\bibfnamefont {J.}~\bibnamefont {Kapusta}}, \
  and\ \bibinfo {author} {\bibfnamefont {C.}~\bibnamefont {Young}},\ }\href
  {\doibase 10.1103/PhysRevC.92.044904} {\bibfield  {journal} {\bibinfo
  {journal} {Phys. Rev.}\ }\textbf {\bibinfo {volume} {C92}},\ \bibinfo {pages}
  {044904} (\bibinfo {year} {2015})},\ \Eprint
  {http://arxiv.org/abs/1506.03408} {arXiv:1506.03408 [nucl-th]} \BibitemShut
  {NoStop}%
%%CITATION = ARXIV:1506.03408;%%
\bibitem [{\citenamefont {Bhattacharyya}\ \emph {et~al.}(2015)\citenamefont
  {Bhattacharyya}, \citenamefont {Ray}, \citenamefont {Samanta},\ and\
  \citenamefont {Sur}}]{Bhattacharyya:2015zka}%
  \BibitemOpen
  \bibfield  {author} {\bibinfo {author} {\bibfnamefont {A.}~\bibnamefont
  {Bhattacharyya}}, \bibinfo {author} {\bibfnamefont {R.}~\bibnamefont {Ray}},
  \bibinfo {author} {\bibfnamefont {S.}~\bibnamefont {Samanta}}, \ and\
  \bibinfo {author} {\bibfnamefont {S.}~\bibnamefont {Sur}},\ }\href {\doibase
  10.1103/PhysRevC.91.041901} {\bibfield  {journal} {\bibinfo  {journal} {Phys.
  Rev.}\ }\textbf {\bibinfo {volume} {C91}},\ \bibinfo {pages} {041901}
  (\bibinfo {year} {2015})},\ \Eprint {http://arxiv.org/abs/1502.00889}
  {arXiv:1502.00889 [hep-ph]} \BibitemShut {NoStop}%
%%CITATION = ARXIV:1502.00889;%%
\bibitem [{\citenamefont {Bhattacharyya}\ \emph {et~al.}(2016)\citenamefont
  {Bhattacharyya}, \citenamefont {Ghosh}, \citenamefont {Ray},\ and\
  \citenamefont {Samanta}}]{Bhattacharyya:2015pra}%
  \BibitemOpen
  \bibfield  {author} {\bibinfo {author} {\bibfnamefont {A.}~\bibnamefont
  {Bhattacharyya}}, \bibinfo {author} {\bibfnamefont {S.~K.}\ \bibnamefont
  {Ghosh}}, \bibinfo {author} {\bibfnamefont {R.}~\bibnamefont {Ray}}, \ and\
  \bibinfo {author} {\bibfnamefont {S.}~\bibnamefont {Samanta}},\ }\href
  {\doibase 10.1209/0295-5075/115/62003} {\bibfield  {journal} {\bibinfo
  {journal} {EPL}\ }\textbf {\bibinfo {volume} {115}},\ \bibinfo {pages}
  {62003} (\bibinfo {year} {2016})},\ \Eprint {http://arxiv.org/abs/1504.04533}
  {arXiv:1504.04533 [hep-ph]} \BibitemShut {NoStop}%
%%CITATION = ARXIV:1504.04533;%%
\bibitem [{\citenamefont {Begun}(2016)}]{Begun:2016cva}%
  \BibitemOpen
  \bibfield  {author} {\bibinfo {author} {\bibfnamefont {V.}~\bibnamefont
  {Begun}},\ }\href {\doibase 10.1103/PhysRevC.94.054904} {\bibfield  {journal}
  {\bibinfo  {journal} {Phys. Rev.}\ }\textbf {\bibinfo {volume} {C94}},\
  \bibinfo {pages} {054904} (\bibinfo {year} {2016})},\ \Eprint
  {http://arxiv.org/abs/1603.02254} {arXiv:1603.02254 [nucl-th]} \BibitemShut
  {NoStop}%
%%CITATION = ARXIV:1603.02254;%%
\bibitem [{\citenamefont {Bhattacharyya}\ \emph {et~al.}(2017)\citenamefont
  {Bhattacharyya}, \citenamefont {Ghosh}, \citenamefont {Maity}, \citenamefont
  {Raha}, \citenamefont {Ray}, \citenamefont {Saha}, \citenamefont {Samanta},\
  and\ \citenamefont {Upadhaya}}]{Bhattacharyya:2017gwt}%
  \BibitemOpen
  \bibfield  {author} {\bibinfo {author} {\bibfnamefont {A.}~\bibnamefont
  {Bhattacharyya}}, \bibinfo {author} {\bibfnamefont {S.~K.}\ \bibnamefont
  {Ghosh}}, \bibinfo {author} {\bibfnamefont {S.}~\bibnamefont {Maity}},
  \bibinfo {author} {\bibfnamefont {S.}~\bibnamefont {Raha}}, \bibinfo {author}
  {\bibfnamefont {R.}~\bibnamefont {Ray}}, \bibinfo {author} {\bibfnamefont
  {K.}~\bibnamefont {Saha}}, \bibinfo {author} {\bibfnamefont {S.}~\bibnamefont
  {Samanta}}, \ and\ \bibinfo {author} {\bibfnamefont {S.}~\bibnamefont
  {Upadhaya}},\ }\href@noop {} {\  (\bibinfo {year} {2017})},\ \Eprint
  {http://arxiv.org/abs/1708.04549} {arXiv:1708.04549 [hep-ph]} \BibitemShut
  {NoStop}%
%%CITATION = ARXIV:1708.04549;%%
\bibitem [{\citenamefont {Andronic}\ \emph {et~al.}(2018)\citenamefont
  {Andronic}, \citenamefont {Braun-Munzinger}, \citenamefont {Redlich},\ and\
  \citenamefont {Stachel}}]{Andronic:2017pug}%
  \BibitemOpen
  \bibfield  {author} {\bibinfo {author} {\bibfnamefont {A.}~\bibnamefont
  {Andronic}}, \bibinfo {author} {\bibfnamefont {P.}~\bibnamefont
  {Braun-Munzinger}}, \bibinfo {author} {\bibfnamefont {K.}~\bibnamefont
  {Redlich}}, \ and\ \bibinfo {author} {\bibfnamefont {J.}~\bibnamefont
  {Stachel}},\ }\href {\doibase 10.1038/s41586-018-0491-6} {\bibfield
  {journal} {\bibinfo  {journal} {Nature}\ }\textbf {\bibinfo {volume} {561}},\
  \bibinfo {pages} {321} (\bibinfo {year} {2018})},\ \Eprint
  {http://arxiv.org/abs/1710.09425} {arXiv:1710.09425 [nucl-th]} \BibitemShut
  {NoStop}%
%%CITATION = ARXIV:1710.09425;%%
\bibitem [{\citenamefont {Ghosh}\ \emph {et~al.}(2018)\citenamefont {Ghosh},
  \citenamefont {Ghosh},\ and\ \citenamefont {Bhattacharyya}}]{Ghosh:2018nqi}%
  \BibitemOpen
  \bibfield  {author} {\bibinfo {author} {\bibfnamefont {S.}~\bibnamefont
  {Ghosh}}, \bibinfo {author} {\bibfnamefont {S.}~\bibnamefont {Ghosh}}, \ and\
  \bibinfo {author} {\bibfnamefont {S.}~\bibnamefont {Bhattacharyya}},\ }\href
  {\doibase 10.1103/PhysRevC.98.045202} {\bibfield  {journal} {\bibinfo
  {journal} {Phys. Rev.}\ }\textbf {\bibinfo {volume} {C98}},\ \bibinfo {pages}
  {045202} (\bibinfo {year} {2018})},\ \Eprint
  {http://arxiv.org/abs/1807.03188} {arXiv:1807.03188 [hep-ph]} \BibitemShut
  {NoStop}%
%%CITATION = ARXIV:1807.03188;%%
\bibitem [{\citenamefont {Dash}\ \emph {et~al.}(2018)\citenamefont {Dash},
  \citenamefont {Samanta},\ and\ \citenamefont {Mohanty}}]{Dash:2018can}%
  \BibitemOpen
  \bibfield  {author} {\bibinfo {author} {\bibfnamefont {A.}~\bibnamefont
  {Dash}}, \bibinfo {author} {\bibfnamefont {S.}~\bibnamefont {Samanta}}, \
  and\ \bibinfo {author} {\bibfnamefont {B.}~\bibnamefont {Mohanty}},\ }\href
  {\doibase 10.1103/PhysRevC.97.055208} {\bibfield  {journal} {\bibinfo
  {journal} {Phys. Rev.}\ }\textbf {\bibinfo {volume} {C97}},\ \bibinfo {pages}
  {055208} (\bibinfo {year} {2018})},\ \Eprint
  {http://arxiv.org/abs/1802.04998} {arXiv:1802.04998 [nucl-th]} \BibitemShut
  {NoStop}%
%%CITATION = ARXIV:1802.04998;%%
\bibitem [{\citenamefont {Biswas}(2020{\natexlab{a}})}]{Biswas:2020dsc}%
  \BibitemOpen
  \bibfield  {author} {\bibinfo {author} {\bibfnamefont {D.}~\bibnamefont
  {Biswas}},\ }\href@noop {} {\  (\bibinfo {year} {2020}{\natexlab{a}})},\
  \Eprint {http://arxiv.org/abs/2003.10425} {arXiv:2003.10425 [hep-ph]}
  \BibitemShut {NoStop}%
\bibitem [{\citenamefont {Biswas}(2020{\natexlab{b}})}]{Biswas:2020kpu}%
  \BibitemOpen
  \bibfield  {author} {\bibinfo {author} {\bibfnamefont {D.}~\bibnamefont
  {Biswas}},\ }\href@noop {} {\  (\bibinfo {year} {2020}{\natexlab{b}})},\
  \Eprint {http://arxiv.org/abs/2007.07680} {arXiv:2007.07680 [nucl-th]}
  \BibitemShut {NoStop}%
\bibitem [{\citenamefont {Alba}\ \emph {et~al.}(2014)\citenamefont {Alba},
  \citenamefont {Alberico}, \citenamefont {Bellwied}, \citenamefont {Bluhm},
  \citenamefont {Mantovani~Sarti}, \citenamefont {Nahrgang},\ and\
  \citenamefont {Ratti}}]{Alba:2014eba}%
  \BibitemOpen
  \bibfield  {author} {\bibinfo {author} {\bibfnamefont {P.}~\bibnamefont
  {Alba}}, \bibinfo {author} {\bibfnamefont {W.}~\bibnamefont {Alberico}},
  \bibinfo {author} {\bibfnamefont {R.}~\bibnamefont {Bellwied}}, \bibinfo
  {author} {\bibfnamefont {M.}~\bibnamefont {Bluhm}}, \bibinfo {author}
  {\bibfnamefont {V.}~\bibnamefont {Mantovani~Sarti}}, \bibinfo {author}
  {\bibfnamefont {M.}~\bibnamefont {Nahrgang}}, \ and\ \bibinfo {author}
  {\bibfnamefont {C.}~\bibnamefont {Ratti}},\ }\href {\doibase
  10.1016/j.physletb.2014.09.052} {\bibfield  {journal} {\bibinfo  {journal}
  {Phys. Lett.}\ }\textbf {\bibinfo {volume} {B738}},\ \bibinfo {pages} {305}
  (\bibinfo {year} {2014})},\ \Eprint {http://arxiv.org/abs/1403.4903}
  {arXiv:1403.4903 [hep-ph]} \BibitemShut {NoStop}%
%%CITATION = ARXIV:1403.4903;%%
\bibitem [{\citenamefont {Chatterjee}\ \emph {et~al.}(2015)\citenamefont
  {Chatterjee}, \citenamefont {Das}, \citenamefont {Kumar}, \citenamefont
  {Mishra}, \citenamefont {Mohanty}, \citenamefont {Sahoo},\ and\ \citenamefont
  {Sharma}}]{Chatterjee:2015fua}%
  \BibitemOpen
  \bibfield  {author} {\bibinfo {author} {\bibfnamefont {S.}~\bibnamefont
  {Chatterjee}}, \bibinfo {author} {\bibfnamefont {S.}~\bibnamefont {Das}},
  \bibinfo {author} {\bibfnamefont {L.}~\bibnamefont {Kumar}}, \bibinfo
  {author} {\bibfnamefont {D.}~\bibnamefont {Mishra}}, \bibinfo {author}
  {\bibfnamefont {B.}~\bibnamefont {Mohanty}}, \bibinfo {author} {\bibfnamefont
  {R.}~\bibnamefont {Sahoo}}, \ and\ \bibinfo {author} {\bibfnamefont
  {N.}~\bibnamefont {Sharma}},\ }\href {\doibase 10.1155/2015/349013}
  {\bibfield  {journal} {\bibinfo  {journal} {Adv. High Energy Phys.}\ }\textbf
  {\bibinfo {volume} {2015}},\ \bibinfo {pages} {349013} (\bibinfo {year}
  {2015})}\BibitemShut {NoStop}%
%%CITATION = 00642,2015,349013;%%
\bibitem [{\citenamefont {Chatterjee}\ \emph {et~al.}(2017)\citenamefont
  {Chatterjee}, \citenamefont {Mishra}, \citenamefont {Mohanty},\ and\
  \citenamefont {Samanta}}]{Chatterjee:2017yhp}%
  \BibitemOpen
  \bibfield  {author} {\bibinfo {author} {\bibfnamefont {S.}~\bibnamefont
  {Chatterjee}}, \bibinfo {author} {\bibfnamefont {D.}~\bibnamefont {Mishra}},
  \bibinfo {author} {\bibfnamefont {B.}~\bibnamefont {Mohanty}}, \ and\
  \bibinfo {author} {\bibfnamefont {S.}~\bibnamefont {Samanta}},\ }\href
  {\doibase 10.1103/PhysRevC.96.054907} {\bibfield  {journal} {\bibinfo
  {journal} {Phys. Rev.}\ }\textbf {\bibinfo {volume} {C96}},\ \bibinfo {pages}
  {054907} (\bibinfo {year} {2017})},\ \Eprint
  {http://arxiv.org/abs/1708.08152} {arXiv:1708.08152 [nucl-th]} \BibitemShut
  {NoStop}%
%%CITATION = ARXIV:1708.08152;%%
\bibitem [{\citenamefont {Adak}\ \emph {et~al.}(2017)\citenamefont {Adak},
  \citenamefont {Das}, \citenamefont {Ghosh}, \citenamefont {Ray},\ and\
  \citenamefont {Samanta}}]{Adak:2016jtk}%
  \BibitemOpen
  \bibfield  {author} {\bibinfo {author} {\bibfnamefont {R.~P.}\ \bibnamefont
  {Adak}}, \bibinfo {author} {\bibfnamefont {S.}~\bibnamefont {Das}}, \bibinfo
  {author} {\bibfnamefont {S.~K.}\ \bibnamefont {Ghosh}}, \bibinfo {author}
  {\bibfnamefont {R.}~\bibnamefont {Ray}}, \ and\ \bibinfo {author}
  {\bibfnamefont {S.}~\bibnamefont {Samanta}},\ }\href {\doibase
  10.1103/PhysRevC.96.014902} {\bibfield  {journal} {\bibinfo  {journal} {Phys.
  Rev.}\ }\textbf {\bibinfo {volume} {C96}},\ \bibinfo {pages} {014902}
  (\bibinfo {year} {2017})},\ \Eprint {http://arxiv.org/abs/1609.05318}
  {arXiv:1609.05318 [nucl-th]} \BibitemShut {NoStop}%
%%CITATION = ARXIV:1609.05318;%%
\bibitem [{\citenamefont {Cleymans}\ \emph {et~al.}(2005)\citenamefont
  {Cleymans}, \citenamefont {Oeschler}, \citenamefont {Redlich},\ and\
  \citenamefont {Wheaton}}]{Cleymans:2004hj}%
  \BibitemOpen
  \bibfield  {author} {\bibinfo {author} {\bibfnamefont {J.}~\bibnamefont
  {Cleymans}}, \bibinfo {author} {\bibfnamefont {H.}~\bibnamefont {Oeschler}},
  \bibinfo {author} {\bibfnamefont {K.}~\bibnamefont {Redlich}}, \ and\
  \bibinfo {author} {\bibfnamefont {S.}~\bibnamefont {Wheaton}},\ }\href
  {\doibase 10.1016/j.physletb.2005.03.074} {\bibfield  {journal} {\bibinfo
  {journal} {Phys. Lett.}\ }\textbf {\bibinfo {volume} {B615}},\ \bibinfo
  {pages} {50} (\bibinfo {year} {2005})},\ \Eprint
  {http://arxiv.org/abs/hep-ph/0411187} {arXiv:hep-ph/0411187 [hep-ph]}
  \BibitemShut {NoStop}%
%%CITATION = HEP-PH/0411187;%%
\bibitem [{\citenamefont {Cleymans}\ \emph {et~al.}(2006)\citenamefont
  {Cleymans}, \citenamefont {Oeschler}, \citenamefont {Redlich},\ and\
  \citenamefont {Wheaton}}]{Cleymans:2005xv}%
  \BibitemOpen
  \bibfield  {author} {\bibinfo {author} {\bibfnamefont {J.}~\bibnamefont
  {Cleymans}}, \bibinfo {author} {\bibfnamefont {H.}~\bibnamefont {Oeschler}},
  \bibinfo {author} {\bibfnamefont {K.}~\bibnamefont {Redlich}}, \ and\
  \bibinfo {author} {\bibfnamefont {S.}~\bibnamefont {Wheaton}},\ }\href
  {\doibase 10.1103/PhysRevC.73.034905} {\bibfield  {journal} {\bibinfo
  {journal} {Phys. Rev.}\ }\textbf {\bibinfo {volume} {C73}},\ \bibinfo {pages}
  {034905} (\bibinfo {year} {2006})},\ \Eprint
  {http://arxiv.org/abs/hep-ph/0511094} {arXiv:hep-ph/0511094 [hep-ph]}
  \BibitemShut {NoStop}%
%%CITATION = HEP-PH/0511094;%%
\bibitem [{\citenamefont {Chatterjee}\ \emph {et~al.}(2013)\citenamefont
  {Chatterjee}, \citenamefont {Godbole},\ and\ \citenamefont
  {Gupta}}]{Chatterjee:2013yga}%
  \BibitemOpen
  \bibfield  {author} {\bibinfo {author} {\bibfnamefont {S.}~\bibnamefont
  {Chatterjee}}, \bibinfo {author} {\bibfnamefont {R.~M.}\ \bibnamefont
  {Godbole}}, \ and\ \bibinfo {author} {\bibfnamefont {S.}~\bibnamefont
  {Gupta}},\ }\href {\doibase 10.1016/j.physletb.2013.11.008} {\bibfield
  {journal} {\bibinfo  {journal} {Phys. Lett.}\ }\textbf {\bibinfo {volume}
  {B727}},\ \bibinfo {pages} {554} (\bibinfo {year} {2013})},\ \Eprint
  {http://arxiv.org/abs/1306.2006} {arXiv:1306.2006 [nucl-th]} \BibitemShut
  {NoStop}%
%%CITATION = ARXIV:1306.2006;%%
\bibitem [{\citenamefont {Wheaton}\ and\ \citenamefont
  {Cleymans}(2009)}]{Wheaton:2004qb}%
  \BibitemOpen
  \bibfield  {author} {\bibinfo {author} {\bibfnamefont {S.}~\bibnamefont
  {Wheaton}}\ and\ \bibinfo {author} {\bibfnamefont {J.}~\bibnamefont
  {Cleymans}},\ }\href {\doibase 10.1016/j.cpc.2008.08.001} {\bibfield
  {journal} {\bibinfo  {journal} {Comput. Phys. Commun.}\ }\textbf {\bibinfo
  {volume} {180}},\ \bibinfo {pages} {84} (\bibinfo {year} {2009})},\ \Eprint
  {http://arxiv.org/abs/hep-ph/0407174} {arXiv:hep-ph/0407174 [hep-ph]}
  \BibitemShut {NoStop}%
%%CITATION = HEP-PH/0407174;%%
\bibitem [{\citenamefont {Torrieri}\ \emph {et~al.}(2005)\citenamefont
  {Torrieri}, \citenamefont {Steinke}, \citenamefont {Broniowski},
  \citenamefont {Florkowski}, \citenamefont {Letessier},\ and\ \citenamefont
  {Rafelski}}]{Torrieri:2004zz}%
  \BibitemOpen
  \bibfield  {author} {\bibinfo {author} {\bibfnamefont {G.}~\bibnamefont
  {Torrieri}}, \bibinfo {author} {\bibfnamefont {S.}~\bibnamefont {Steinke}},
  \bibinfo {author} {\bibfnamefont {W.}~\bibnamefont {Broniowski}}, \bibinfo
  {author} {\bibfnamefont {W.}~\bibnamefont {Florkowski}}, \bibinfo {author}
  {\bibfnamefont {J.}~\bibnamefont {Letessier}}, \ and\ \bibinfo {author}
  {\bibfnamefont {J.}~\bibnamefont {Rafelski}},\ }\href {\doibase
  10.1016/j.cpc.2005.01.004} {\bibfield  {journal} {\bibinfo  {journal}
  {Comput. Phys. Commun.}\ }\textbf {\bibinfo {volume} {167}},\ \bibinfo
  {pages} {229} (\bibinfo {year} {2005})},\ \Eprint
  {http://arxiv.org/abs/nucl-th/0404083} {arXiv:nucl-th/0404083 [nucl-th]}
  \BibitemShut {NoStop}%
%%CITATION = NUCL-TH/0404083;%%
\bibitem [{\citenamefont {Kisiel}\ \emph {et~al.}(2006)\citenamefont {Kisiel},
  \citenamefont {Taluc}, \citenamefont {Broniowski},\ and\ \citenamefont
  {Florkowski}}]{Kisiel:2005hn}%
  \BibitemOpen
  \bibfield  {author} {\bibinfo {author} {\bibfnamefont {A.}~\bibnamefont
  {Kisiel}}, \bibinfo {author} {\bibfnamefont {T.}~\bibnamefont {Taluc}},
  \bibinfo {author} {\bibfnamefont {W.}~\bibnamefont {Broniowski}}, \ and\
  \bibinfo {author} {\bibfnamefont {W.}~\bibnamefont {Florkowski}},\ }\href
  {\doibase 10.1016/j.cpc.2005.11.010} {\bibfield  {journal} {\bibinfo
  {journal} {Comput. Phys. Commun.}\ }\textbf {\bibinfo {volume} {174}},\
  \bibinfo {pages} {669} (\bibinfo {year} {2006})},\ \Eprint
  {http://arxiv.org/abs/nucl-th/0504047} {arXiv:nucl-th/0504047 [nucl-th]}
  \BibitemShut {NoStop}%
%%CITATION = NUCL-TH/0504047;%%
\bibitem [{\citenamefont {Adamczyk}\ \emph {et~al.}(2017)\citenamefont
  {Adamczyk} \emph {et~al.}}]{Adamczyk:2017iwn}%
  \BibitemOpen
  \bibfield  {author} {\bibinfo {author} {\bibfnamefont {L.}~\bibnamefont
  {Adamczyk}} \emph {et~al.} (\bibinfo {collaboration} {STAR}),\ }\href
  {\doibase 10.1103/PhysRevC.96.044904} {\bibfield  {journal} {\bibinfo
  {journal} {Phys. Rev.}\ }\textbf {\bibinfo {volume} {C96}},\ \bibinfo {pages}
  {044904} (\bibinfo {year} {2017})},\ \Eprint
  {http://arxiv.org/abs/1701.07065} {arXiv:1701.07065 [nucl-ex]} \BibitemShut
  {NoStop}%
%%CITATION = ARXIV:1701.07065;%%
\bibitem [{\citenamefont {Bhattacharyya}\ \emph {et~al.}(2020)\citenamefont
  {Bhattacharyya}, \citenamefont {Biswas}, \citenamefont {Ghosh}, \citenamefont
  {Ray},\ and\ \citenamefont {Singha}}]{Bhattacharyya:2019cer}%
  \BibitemOpen
  \bibfield  {author} {\bibinfo {author} {\bibfnamefont {S.}~\bibnamefont
  {Bhattacharyya}}, \bibinfo {author} {\bibfnamefont {D.}~\bibnamefont
  {Biswas}}, \bibinfo {author} {\bibfnamefont {S.~K.}\ \bibnamefont {Ghosh}},
  \bibinfo {author} {\bibfnamefont {R.}~\bibnamefont {Ray}}, \ and\ \bibinfo
  {author} {\bibfnamefont {P.}~\bibnamefont {Singha}},\ }\href {\doibase
  10.1103/PhysRevD.101.054002} {\bibfield  {journal} {\bibinfo  {journal}
  {Phys. Rev. D}\ }\textbf {\bibinfo {volume} {101}},\ \bibinfo {pages}
  {054002} (\bibinfo {year} {2020})},\ \Eprint
  {http://arxiv.org/abs/1911.04828} {arXiv:1911.04828 [hep-ph]} \BibitemShut
  {NoStop}%
\bibitem [{\citenamefont {Becattini}(2007)}]{Becattini:2007wt}%
  \BibitemOpen
  \bibfield  {author} {\bibinfo {author} {\bibfnamefont {F.}~\bibnamefont
  {Becattini}},\ }\href@noop {} {\  (\bibinfo {year} {2007})},\ \Eprint
  {http://arxiv.org/abs/0707.4154} {arXiv:0707.4154 [nucl-th]} \BibitemShut
  {NoStop}%
%%CITATION = ARXIV:0707.4154;%%
\bibitem [{\citenamefont {Cayley}(1889)}]{Cayley:1889}%
  \BibitemOpen
  \bibfield  {author} {\bibinfo {author} {\bibfnamefont {A.}~\bibnamefont
  {Cayley}},\ }\href@noop {} {\bibfield  {journal} {\bibinfo  {journal} {Quart.
  J. Pure Appl. Math.}\ }\textbf {\bibinfo {volume} {23}},\ \bibinfo {pages}
  {376–378} (\bibinfo {year} {1889})}\BibitemShut {NoStop}%
\bibitem [{\citenamefont {Cayley}()}]{Cayley:1897}%
  \BibitemOpen
  \bibfield  {author} {\bibinfo {author} {\bibfnamefont {A.}~\bibnamefont
  {Cayley}},\ }\href@noop {} {\emph {\bibinfo {title} {{The collected
  mathematical papers. {V}olume~13}}}},\ Cambridge Library Collection\
  (\bibinfo  {publisher} {Cambridge University Press})\BibitemShut {NoStop}%
\bibitem [{\citenamefont {Graham}\ and\ \citenamefont {Hell}()}]{Graham1985}%
  \BibitemOpen
  \bibfield  {author} {\bibinfo {author} {\bibfnamefont {R.~L.}\ \bibnamefont
  {Graham}}\ and\ \bibinfo {author} {\bibfnamefont {P.}~\bibnamefont {Hell}},\
  }\href {\doibase 10.1109/MAHC.1985.10011} {\bibfield  {journal} {\bibinfo
  {journal} {Ann. Hist. Comput.}\ }\textbf {\bibinfo {volume} {7}},\
  10.1109/MAHC.1985.10011}\BibitemShut {NoStop}%
\bibitem [{\citenamefont {Cormen}\ \emph {et~al.}(2009)\citenamefont {Cormen},
  \citenamefont {Leiserson}, \citenamefont {Rivest},\ and\ \citenamefont
  {Stein}}]{Cormen2009}%
  \BibitemOpen
  \bibfield  {author} {\bibinfo {author} {\bibfnamefont {T.~H.}\ \bibnamefont
  {Cormen}}, \bibinfo {author} {\bibfnamefont {C.~E.}\ \bibnamefont
  {Leiserson}}, \bibinfo {author} {\bibfnamefont {R.~L.}\ \bibnamefont
  {Rivest}}, \ and\ \bibinfo {author} {\bibfnamefont {C.}~\bibnamefont
  {Stein}},\ }\href@noop {} {\emph {\bibinfo {title} {{Introduction to
  algorithms}}}},\ \bibinfo {edition} {3rd}\ ed.\ (\bibinfo  {publisher} {MIT
  Press, Cambridge, MA},\ \bibinfo {year} {2009})\ pp.\ \bibinfo {pages}
  {xx+1292}\BibitemShut {NoStop}%
\bibitem [{\citenamefont {Chakraborty}\ \emph {et~al.}(2019)\citenamefont
  {Chakraborty}, \citenamefont {Chowdhury}, \citenamefont {Chakraborty},
  \citenamefont {Mehera},\ and\ \citenamefont {Pal}}]{Chakraborty:2019algo}%
  \BibitemOpen
  \bibfield  {author} {\bibinfo {author} {\bibfnamefont {M.}~\bibnamefont
  {Chakraborty}}, \bibinfo {author} {\bibfnamefont {S.}~\bibnamefont
  {Chowdhury}}, \bibinfo {author} {\bibfnamefont {J.}~\bibnamefont
  {Chakraborty}}, \bibinfo {author} {\bibfnamefont {R.}~\bibnamefont {Mehera}},
  \ and\ \bibinfo {author} {\bibfnamefont {R.~K.}\ \bibnamefont {Pal}},\ }\href
  {\doibase 10.1007/s40747-018-0079-7} {\bibfield  {journal} {\bibinfo
  {journal} {Complex \& Intelligent Systems}\ }\textbf {\bibinfo {volume}
  {5}},\ \bibinfo {pages} {265} (\bibinfo {year} {2019})}\BibitemShut {NoStop}%
\bibitem [{\citenamefont {Kumar}(2013)}]{Kumar:2012fb}%
  \BibitemOpen
  \bibfield  {author} {\bibinfo {author} {\bibfnamefont {L.}~\bibnamefont
  {Kumar}} (\bibinfo {collaboration} {STAR}),\ }\href {\doibase
  10.1016/j.nuclphysa.2013.01.070} {\bibfield  {journal} {\bibinfo  {journal}
  {Nucl. Phys.}\ }\textbf {\bibinfo {volume} {A904-905}},\ \bibinfo {pages}
  {256c} (\bibinfo {year} {2013})},\ \Eprint {http://arxiv.org/abs/1211.1350}
  {arXiv:1211.1350 [nucl-ex]} \BibitemShut {NoStop}%
%%CITATION = ARXIV:1211.1350;%%
\bibitem [{\citenamefont {Das}(2013)}]{Das:2012yq}%
  \BibitemOpen
  \bibfield  {author} {\bibinfo {author} {\bibfnamefont {S.}~\bibnamefont
  {Das}} (\bibinfo {collaboration} {STAR}),\ }\href {\doibase
  10.1016/j.nuclphysa.2013.02.158} {\bibfield  {journal} {\bibinfo  {journal}
  {Nucl. Phys.}\ }\textbf {\bibinfo {volume} {A904-905}},\ \bibinfo {pages}
  {891c} (\bibinfo {year} {2013})},\ \Eprint {http://arxiv.org/abs/1210.6099}
  {arXiv:1210.6099 [nucl-ex]} \BibitemShut {NoStop}%
%%CITATION = ARXIV:1210.6099;%%
\bibitem [{\citenamefont {Adler}\ \emph
  {et~al.}(2002{\natexlab{a}})\citenamefont {Adler} \emph
  {et~al.}}]{Adler:2002uv}%
  \BibitemOpen
  \bibfield  {author} {\bibinfo {author} {\bibfnamefont {C.}~\bibnamefont
  {Adler}} \emph {et~al.} (\bibinfo {collaboration} {STAR}),\ }\href {\doibase
  10.1103/PhysRevLett.89.092301} {\bibfield  {journal} {\bibinfo  {journal}
  {Phys. Rev. Lett.}\ }\textbf {\bibinfo {volume} {89}},\ \bibinfo {pages}
  {092301} (\bibinfo {year} {2002}{\natexlab{a}})},\ \Eprint
  {http://arxiv.org/abs/nucl-ex/0203016} {arXiv:nucl-ex/0203016 [nucl-ex]}
  \BibitemShut {NoStop}%
%%CITATION = NUCL-EX/0203016;%%
\bibitem [{\citenamefont {Adams}\ \emph {et~al.}(2004)\citenamefont {Adams}
  \emph {et~al.}}]{Adams:2003fy}%
  \BibitemOpen
  \bibfield  {author} {\bibinfo {author} {\bibfnamefont {J.}~\bibnamefont
  {Adams}} \emph {et~al.} (\bibinfo {collaboration} {STAR}),\ }\href {\doibase
  10.1103/PhysRevLett.92.182301} {\bibfield  {journal} {\bibinfo  {journal}
  {Phys. Rev. Lett.}\ }\textbf {\bibinfo {volume} {92}},\ \bibinfo {pages}
  {182301} (\bibinfo {year} {2004})},\ \Eprint
  {http://arxiv.org/abs/nucl-ex/0307024} {arXiv:nucl-ex/0307024 [nucl-ex]}
  \BibitemShut {NoStop}%
%%CITATION = NUCL-EX/0307024;%%
\bibitem [{\citenamefont {Zhu}(2012)}]{Zhu:2012ph}%
  \BibitemOpen
  \bibfield  {author} {\bibinfo {author} {\bibfnamefont {X.}~\bibnamefont
  {Zhu}} (\bibinfo {collaboration} {STAR}),\ }\href {\doibase
  10.5506/APhysPolBSupp.5.213} {\bibfield  {journal} {\bibinfo  {journal} {Acta
  Phys. Polon. Supp.}\ }\textbf {\bibinfo {volume} {5}},\ \bibinfo {pages}
  {213} (\bibinfo {year} {2012})},\ \Eprint {http://arxiv.org/abs/1203.5183}
  {arXiv:1203.5183 [nucl-ex]} \BibitemShut {NoStop}%
%%CITATION = ARXIV:1203.5183;%%
\bibitem [{\citenamefont {Zhao}(2014)}]{Zhao:2014mva}%
  \BibitemOpen
  \bibfield  {author} {\bibinfo {author} {\bibfnamefont {F.}~\bibnamefont
  {Zhao}} (\bibinfo {collaboration} {STAR}),\ }\href {\doibase
  10.1088/1742-6596/509/1/012085} {\bibfield  {journal} {\bibinfo  {journal}
  {J. Phys. Conf. Ser.}\ }\textbf {\bibinfo {volume} {509}},\ \bibinfo {pages}
  {012085} (\bibinfo {year} {2014})}\BibitemShut {NoStop}%
%%CITATION = 00462,509,012085;%%
\bibitem [{\citenamefont {Kumar}(2014)}]{Kumar:2014tca}%
  \BibitemOpen
  \bibfield  {author} {\bibinfo {author} {\bibfnamefont {L.}~\bibnamefont
  {Kumar}} (\bibinfo {collaboration} {STAR}),\ }\href {\doibase
  10.1016/j.nuclphysa.2014.08.085} {\bibfield  {journal} {\bibinfo  {journal}
  {Nucl. Phys.}\ }\textbf {\bibinfo {volume} {A931}},\ \bibinfo {pages} {1114}
  (\bibinfo {year} {2014})},\ \Eprint {http://arxiv.org/abs/1408.4209}
  {arXiv:1408.4209 [nucl-ex]} \BibitemShut {NoStop}%
%%CITATION = ARXIV:1408.4209;%%
\bibitem [{\citenamefont {Das}(2014)}]{Das:2014kja}%
  \BibitemOpen
  \bibfield  {author} {\bibinfo {author} {\bibfnamefont {S.}~\bibnamefont
  {Das}} (\bibinfo {collaboration} {STAR}),\ }\href {\doibase
  10.1088/1742-6596/509/1/012066} {\bibfield  {journal} {\bibinfo  {journal}
  {J. Phys. Conf. Ser.}\ }\textbf {\bibinfo {volume} {509}},\ \bibinfo {pages}
  {012066} (\bibinfo {year} {2014})},\ \Eprint {http://arxiv.org/abs/1402.0255}
  {arXiv:1402.0255 [nucl-ex]} \BibitemShut {NoStop}%
%%CITATION = ARXIV:1402.0255;%%
\bibitem [{\citenamefont {Abelev}\ \emph
  {et~al.}(2009{\natexlab{a}})\citenamefont {Abelev} \emph
  {et~al.}}]{Abelev:2008ab}%
  \BibitemOpen
  \bibfield  {author} {\bibinfo {author} {\bibfnamefont {B.~I.}\ \bibnamefont
  {Abelev}} \emph {et~al.} (\bibinfo {collaboration} {STAR}),\ }\href {\doibase
  10.1103/PhysRevC.79.034909} {\bibfield  {journal} {\bibinfo  {journal} {Phys.
  Rev.}\ }\textbf {\bibinfo {volume} {C79}},\ \bibinfo {pages} {034909}
  (\bibinfo {year} {2009}{\natexlab{a}})},\ \Eprint
  {http://arxiv.org/abs/0808.2041} {arXiv:0808.2041 [nucl-ex]} \BibitemShut
  {NoStop}%
%%CITATION = ARXIV:0808.2041;%%
\bibitem [{\citenamefont {Aggarwal}\ \emph {et~al.}(2011)\citenamefont
  {Aggarwal} \emph {et~al.}}]{Aggarwal:2010ig}%
  \BibitemOpen
  \bibfield  {author} {\bibinfo {author} {\bibfnamefont {M.~M.}\ \bibnamefont
  {Aggarwal}} \emph {et~al.} (\bibinfo {collaboration} {STAR}),\ }\href
  {\doibase 10.1103/PhysRevC.83.024901} {\bibfield  {journal} {\bibinfo
  {journal} {Phys. Rev.}\ }\textbf {\bibinfo {volume} {C83}},\ \bibinfo {pages}
  {024901} (\bibinfo {year} {2011})},\ \Eprint {http://arxiv.org/abs/1010.0142}
  {arXiv:1010.0142 [nucl-ex]} \BibitemShut {NoStop}%
%%CITATION = ARXIV:1010.0142;%%
\bibitem [{\citenamefont {Abelev}\ \emph
  {et~al.}(2009{\natexlab{b}})\citenamefont {Abelev} \emph
  {et~al.}}]{Abelev:2008aa}%
  \BibitemOpen
  \bibfield  {author} {\bibinfo {author} {\bibfnamefont {B.~I.}\ \bibnamefont
  {Abelev}} \emph {et~al.} (\bibinfo {collaboration} {STAR}),\ }\href {\doibase
  10.1103/PhysRevC.79.064903} {\bibfield  {journal} {\bibinfo  {journal} {Phys.
  Rev.}\ }\textbf {\bibinfo {volume} {C79}},\ \bibinfo {pages} {064903}
  (\bibinfo {year} {2009}{\natexlab{b}})},\ \Eprint
  {http://arxiv.org/abs/0809.4737} {arXiv:0809.4737 [nucl-ex]} \BibitemShut
  {NoStop}%
%%CITATION = ARXIV:0809.4737;%%
\bibitem [{\citenamefont {Adcox}\ \emph {et~al.}(2002)\citenamefont {Adcox}
  \emph {et~al.}}]{Adcox:2002au}%
  \BibitemOpen
  \bibfield  {author} {\bibinfo {author} {\bibfnamefont {K.}~\bibnamefont
  {Adcox}} \emph {et~al.} (\bibinfo {collaboration} {PHENIX}),\ }\href
  {\doibase 10.1103/PhysRevLett.89.092302} {\bibfield  {journal} {\bibinfo
  {journal} {Phys. Rev. Lett.}\ }\textbf {\bibinfo {volume} {89}},\ \bibinfo
  {pages} {092302} (\bibinfo {year} {2002})},\ \Eprint
  {http://arxiv.org/abs/nucl-ex/0204007} {arXiv:nucl-ex/0204007 [nucl-ex]}
  \BibitemShut {NoStop}%
%%CITATION = NUCL-EX/0204007;%%
\bibitem [{\citenamefont {Adler}\ \emph
  {et~al.}(2002{\natexlab{b}})\citenamefont {Adler} \emph
  {et~al.}}]{Adler:2002xv}%
  \BibitemOpen
  \bibfield  {author} {\bibinfo {author} {\bibfnamefont {C.}~\bibnamefont
  {Adler}} \emph {et~al.} (\bibinfo {collaboration} {STAR}),\ }\href {\doibase
  10.1103/PhysRevC.65.041901} {\bibfield  {journal} {\bibinfo  {journal} {Phys.
  Rev.}\ }\textbf {\bibinfo {volume} {C65}},\ \bibinfo {pages} {041901}
  (\bibinfo {year} {2002}{\natexlab{b}})}\BibitemShut {NoStop}%
%%CITATION = PHRVA,C65,041901;%%
\bibitem [{\citenamefont {Adams}\ \emph {et~al.}(2007)\citenamefont {Adams}
  \emph {et~al.}}]{Adams:2006ke}%
  \BibitemOpen
  \bibfield  {author} {\bibinfo {author} {\bibfnamefont {J.}~\bibnamefont
  {Adams}} \emph {et~al.} (\bibinfo {collaboration} {STAR}),\ }\href {\doibase
  10.1103/PhysRevLett.98.062301} {\bibfield  {journal} {\bibinfo  {journal}
  {Phys. Rev. Lett.}\ }\textbf {\bibinfo {volume} {98}},\ \bibinfo {pages}
  {062301} (\bibinfo {year} {2007})},\ \Eprint
  {http://arxiv.org/abs/nucl-ex/0606014} {arXiv:nucl-ex/0606014 [nucl-ex]}
  \BibitemShut {NoStop}%
%%CITATION = NUCL-EX/0606014;%%
\bibitem [{\citenamefont {Adams}\ \emph {et~al.}(2005)\citenamefont {Adams}
  \emph {et~al.}}]{Adams:2004ux}%
  \BibitemOpen
  \bibfield  {author} {\bibinfo {author} {\bibfnamefont {J.}~\bibnamefont
  {Adams}} \emph {et~al.} (\bibinfo {collaboration} {STAR}),\ }\href {\doibase
  10.1016/j.physletb.2004.12.082} {\bibfield  {journal} {\bibinfo  {journal}
  {Phys. Lett.}\ }\textbf {\bibinfo {volume} {B612}},\ \bibinfo {pages} {181}
  (\bibinfo {year} {2005})},\ \Eprint {http://arxiv.org/abs/nucl-ex/0406003}
  {arXiv:nucl-ex/0406003 [nucl-ex]} \BibitemShut {NoStop}%
%%CITATION = NUCL-EX/0406003;%%
\bibitem [{\citenamefont {Kumar}(2012)}]{Kumar:2012np}%
  \BibitemOpen
  \bibfield  {author} {\bibinfo {author} {\bibfnamefont {L.}~\bibnamefont
  {Kumar}} (\bibinfo {collaboration} {STAR}),\ }\href {\doibase
  10.2478/s11534-012-0097-9} {\bibfield  {journal} {\bibinfo  {journal}
  {Central Eur. J. Phys.}\ }\textbf {\bibinfo {volume} {10}},\ \bibinfo {pages}
  {1274} (\bibinfo {year} {2012})},\ \Eprint {http://arxiv.org/abs/1201.4203}
  {arXiv:1201.4203 [nucl-ex]} \BibitemShut {NoStop}%
%%CITATION = ARXIV:1201.4203;%%
\bibitem [{\citenamefont {Abelev}\ \emph {et~al.}(2012)\citenamefont {Abelev}
  \emph {et~al.}}]{Abelev:2012wca}%
  \BibitemOpen
  \bibfield  {author} {\bibinfo {author} {\bibfnamefont {B.}~\bibnamefont
  {Abelev}} \emph {et~al.} (\bibinfo {collaboration} {ALICE}),\ }\href
  {\doibase 10.1103/PhysRevLett.109.252301} {\bibfield  {journal} {\bibinfo
  {journal} {Phys. Rev. Lett.}\ }\textbf {\bibinfo {volume} {109}},\ \bibinfo
  {pages} {252301} (\bibinfo {year} {2012})},\ \Eprint
  {http://arxiv.org/abs/1208.1974} {arXiv:1208.1974 [hep-ex]} \BibitemShut
  {NoStop}%
%%CITATION = ARXIV:1208.1974;%%
\bibitem [{\citenamefont {Abelev}\ \emph
  {et~al.}(2013{\natexlab{a}})\citenamefont {Abelev} \emph
  {et~al.}}]{Abelev:2013xaa}%
  \BibitemOpen
  \bibfield  {author} {\bibinfo {author} {\bibfnamefont {B.~B.}\ \bibnamefont
  {Abelev}} \emph {et~al.} (\bibinfo {collaboration} {ALICE}),\ }\href
  {\doibase 10.1103/PhysRevLett.111.222301} {\bibfield  {journal} {\bibinfo
  {journal} {Phys. Rev. Lett.}\ }\textbf {\bibinfo {volume} {111}},\ \bibinfo
  {pages} {222301} (\bibinfo {year} {2013}{\natexlab{a}})},\ \Eprint
  {http://arxiv.org/abs/1307.5530} {arXiv:1307.5530 [nucl-ex]} \BibitemShut
  {NoStop}%
%%CITATION = ARXIV:1307.5530;%%
\bibitem [{\citenamefont {Abelev}\ \emph {et~al.}(2014)\citenamefont {Abelev}
  \emph {et~al.}}]{ABELEV:2013zaa}%
  \BibitemOpen
  \bibfield  {author} {\bibinfo {author} {\bibfnamefont {B.~B.}\ \bibnamefont
  {Abelev}} \emph {et~al.} (\bibinfo {collaboration} {ALICE}),\ }\href
  {\doibase 10.1016/j.physletb.2014.05.052, 10.1016/j.physletb.2013.11.048}
  {\bibfield  {journal} {\bibinfo  {journal} {Phys. Lett.}\ }\textbf {\bibinfo
  {volume} {B728}},\ \bibinfo {pages} {216} (\bibinfo {year} {2014})},\
  \bibinfo {note} {[Erratum: Phys. Lett.B734,409(2014)]},\ \Eprint
  {http://arxiv.org/abs/1307.5543} {arXiv:1307.5543 [nucl-ex]} \BibitemShut
  {NoStop}%
%%CITATION = ARXIV:1307.5543;%%
\bibitem [{\citenamefont {Abelev}\ \emph
  {et~al.}(2013{\natexlab{b}})\citenamefont {Abelev} \emph
  {et~al.}}]{Abelev:2013vea}%
  \BibitemOpen
  \bibfield  {author} {\bibinfo {author} {\bibfnamefont {B.}~\bibnamefont
  {Abelev}} \emph {et~al.} (\bibinfo {collaboration} {ALICE}),\ }\href
  {\doibase 10.1103/PhysRevC.88.044910} {\bibfield  {journal} {\bibinfo
  {journal} {Phys. Rev.}\ }\textbf {\bibinfo {volume} {C88}},\ \bibinfo {pages}
  {044910} (\bibinfo {year} {2013}{\natexlab{b}})},\ \Eprint
  {http://arxiv.org/abs/1303.0737} {arXiv:1303.0737 [hep-ex]} \BibitemShut
  {NoStop}%
%%CITATION = ARXIV:1303.0737;%%
\bibitem [{\citenamefont {Tanabashi}\ \emph {et~al.}(2018)\citenamefont
  {Tanabashi} \emph {et~al.}}]{Tanabashi:2018oca}%
  \BibitemOpen
  \bibfield  {author} {\bibinfo {author} {\bibfnamefont {M.}~\bibnamefont
  {Tanabashi}} \emph {et~al.} (\bibinfo {collaboration} {Particle Data
  Group}),\ }\href {\doibase 10.1103/PhysRevD.98.030001} {\bibfield  {journal}
  {\bibinfo  {journal} {Phys. Rev.}\ }\textbf {\bibinfo {volume} {D98}},\
  \bibinfo {pages} {030001} (\bibinfo {year} {2018})}\BibitemShut {NoStop}%
%%CITATION = PHRVA,D98,030001;%%
\bibitem [{\citenamefont {Zschiesche}\ \emph {et~al.}(2002)\citenamefont
  {Zschiesche}, \citenamefont {Schramm}, \citenamefont {Schaffner-Bielich},
  \citenamefont {Stoecker},\ and\ \citenamefont {Greiner}}]{Zschiesche:2002zr}%
  \BibitemOpen
  \bibfield  {author} {\bibinfo {author} {\bibfnamefont {D.}~\bibnamefont
  {Zschiesche}}, \bibinfo {author} {\bibfnamefont {S.}~\bibnamefont {Schramm}},
  \bibinfo {author} {\bibfnamefont {J.}~\bibnamefont {Schaffner-Bielich}},
  \bibinfo {author} {\bibfnamefont {H.}~\bibnamefont {Stoecker}}, \ and\
  \bibinfo {author} {\bibfnamefont {W.}~\bibnamefont {Greiner}},\ }\href
  {\doibase 10.1016/S0370-2693(02)02736-3} {\bibfield  {journal} {\bibinfo
  {journal} {Phys. Lett. B}\ }\textbf {\bibinfo {volume} {547}},\ \bibinfo
  {pages} {7} (\bibinfo {year} {2002})},\ \Eprint
  {http://arxiv.org/abs/nucl-th/0209022} {arXiv:nucl-th/0209022} \BibitemShut
  {NoStop}%
\bibitem [{\citenamefont {Koch}(2004)}]{Koch:2003pj}%
  \BibitemOpen
  \bibfield  {author} {\bibinfo {author} {\bibfnamefont {V.}~\bibnamefont
  {Koch}},\ }\href {\doibase 10.1088/0954-3899/30/1/003} {\bibfield  {journal}
  {\bibinfo  {journal} {J. Phys. G}\ }\textbf {\bibinfo {volume} {30}},\
  \bibinfo {pages} {S41} (\bibinfo {year} {2004})},\ \Eprint
  {http://arxiv.org/abs/nucl-th/0306037} {arXiv:nucl-th/0306037} \BibitemShut
  {NoStop}%
\bibitem [{\citenamefont {Brown}\ and\ \citenamefont
  {Rho}(2002)}]{Brown:2001nh}%
  \BibitemOpen
  \bibfield  {author} {\bibinfo {author} {\bibfnamefont {G.}~\bibnamefont
  {Brown}}\ and\ \bibinfo {author} {\bibfnamefont {M.}~\bibnamefont {Rho}},\
  }\href {\doibase 10.1016/S0370-1573(01)00084-9} {\bibfield  {journal}
  {\bibinfo  {journal} {Phys. Rept.}\ }\textbf {\bibinfo {volume} {363}},\
  \bibinfo {pages} {85} (\bibinfo {year} {2002})},\ \Eprint
  {http://arxiv.org/abs/hep-ph/0103102} {arXiv:hep-ph/0103102} \BibitemShut
  {NoStop}%
\bibitem [{\citenamefont {Hagedorn}(1965)}]{Hagedorn:1965st}%
  \BibitemOpen
  \bibfield  {author} {\bibinfo {author} {\bibfnamefont {R.}~\bibnamefont
  {Hagedorn}},\ }\href@noop {} {\bibfield  {journal} {\bibinfo  {journal}
  {Nuovo Cim. Suppl.}\ }\textbf {\bibinfo {volume} {3}},\ \bibinfo {pages}
  {147} (\bibinfo {year} {1965})}\BibitemShut {NoStop}%
%%CITATION = NUCUA,3,147;%%
\bibitem [{\citenamefont {Witten}(1984)}]{Witten:1984rs}%
  \BibitemOpen
  \bibfield  {author} {\bibinfo {author} {\bibfnamefont {E.}~\bibnamefont
  {Witten}},\ }\href {\doibase 10.1103/PhysRevD.30.272} {\bibfield  {journal}
  {\bibinfo  {journal} {Phys. Rev.}\ }\textbf {\bibinfo {volume} {D30}},\
  \bibinfo {pages} {272} (\bibinfo {year} {1984})}\BibitemShut {NoStop}%
%%CITATION = PHRVA,D30,272;%%
\bibitem [{\citenamefont {Stachel}\ \emph {et~al.}(2014)\citenamefont
  {Stachel}, \citenamefont {Andronic}, \citenamefont {Braun-Munzinger},\ and\
  \citenamefont {Redlich}}]{Stachel:2013zma}%
  \BibitemOpen
  \bibfield  {author} {\bibinfo {author} {\bibfnamefont {J.}~\bibnamefont
  {Stachel}}, \bibinfo {author} {\bibfnamefont {A.}~\bibnamefont {Andronic}},
  \bibinfo {author} {\bibfnamefont {P.}~\bibnamefont {Braun-Munzinger}}, \ and\
  \bibinfo {author} {\bibfnamefont {K.}~\bibnamefont {Redlich}},\ }\href
  {\doibase 10.1088/1742-6596/509/1/012019} {\bibfield  {journal} {\bibinfo
  {journal} {J. Phys. Conf. Ser.}\ }\textbf {\bibinfo {volume} {509}},\
  \bibinfo {pages} {012019} (\bibinfo {year} {2014})},\ \Eprint
  {http://arxiv.org/abs/1311.4662} {arXiv:1311.4662 [nucl-th]} \BibitemShut
  {NoStop}%
\bibitem [{\citenamefont {Huovinen}\ and\ \citenamefont
  {Petreczky}(2018{\natexlab{a}})}]{Huovinen:2017ogf}%
  \BibitemOpen
  \bibfield  {author} {\bibinfo {author} {\bibfnamefont {P.}~\bibnamefont
  {Huovinen}}\ and\ \bibinfo {author} {\bibfnamefont {P.}~\bibnamefont
  {Petreczky}},\ }\href {\doibase 10.1016/j.physletb.2017.12.001} {\bibfield
  {journal} {\bibinfo  {journal} {Phys. Lett. B}\ }\textbf {\bibinfo {volume}
  {777}},\ \bibinfo {pages} {125} (\bibinfo {year} {2018}{\natexlab{a}})},\
  \Eprint {http://arxiv.org/abs/1708.00879} {arXiv:1708.00879 [hep-ph]}
  \BibitemShut {NoStop}%
\bibitem [{\citenamefont {Huovinen}\ and\ \citenamefont
  {Petreczky}(2018{\natexlab{b}})}]{Huovinen:2018ziu}%
  \BibitemOpen
  \bibfield  {author} {\bibinfo {author} {\bibfnamefont {P.}~\bibnamefont
  {Huovinen}}\ and\ \bibinfo {author} {\bibfnamefont {P.}~\bibnamefont
  {Petreczky}},\ }\href {\doibase 10.22323/1.336.0145} {\bibfield  {journal}
  {\bibinfo  {journal} {PoS}\ }\textbf {\bibinfo {volume} {Confinement2018}},\
  \bibinfo {pages} {145} (\bibinfo {year} {2018}{\natexlab{b}})},\ \Eprint
  {http://arxiv.org/abs/1811.09330} {arXiv:1811.09330 [nucl-th]} \BibitemShut
  {NoStop}%
\bibitem [{\citenamefont {Steinheimer}\ \emph {et~al.}(2013)\citenamefont
  {Steinheimer}, \citenamefont {Aichelin},\ and\ \citenamefont
  {Bleicher}}]{Steinheimer:2012rd}%
  \BibitemOpen
  \bibfield  {author} {\bibinfo {author} {\bibfnamefont {J.}~\bibnamefont
  {Steinheimer}}, \bibinfo {author} {\bibfnamefont {J.}~\bibnamefont
  {Aichelin}}, \ and\ \bibinfo {author} {\bibfnamefont {M.}~\bibnamefont
  {Bleicher}},\ }\href {\doibase 10.1103/PhysRevLett.110.042501} {\bibfield
  {journal} {\bibinfo  {journal} {Phys. Rev. Lett.}\ }\textbf {\bibinfo
  {volume} {110}},\ \bibinfo {pages} {042501} (\bibinfo {year} {2013})},\
  \Eprint {http://arxiv.org/abs/1203.5302} {arXiv:1203.5302 [nucl-th]}
  \BibitemShut {NoStop}%
\bibitem [{\citenamefont {Petr\'an}\ \emph {et~al.}(2013)\citenamefont
  {Petr\'an}, \citenamefont {Letessier}, \citenamefont {Petr\'a\v{c}ek},\ and\
  \citenamefont {Rafelski}}]{Petran:2013lja}%
  \BibitemOpen
  \bibfield  {author} {\bibinfo {author} {\bibfnamefont {M.}~\bibnamefont
  {Petr\'an}}, \bibinfo {author} {\bibfnamefont {J.}~\bibnamefont {Letessier}},
  \bibinfo {author} {\bibfnamefont {V.}~\bibnamefont {Petr\'a\v{c}ek}}, \ and\
  \bibinfo {author} {\bibfnamefont {J.}~\bibnamefont {Rafelski}},\ }\href
  {\doibase 10.1103/PhysRevC.88.034907} {\bibfield  {journal} {\bibinfo
  {journal} {Phys. Rev. C}\ }\textbf {\bibinfo {volume} {88}},\ \bibinfo
  {pages} {034907} (\bibinfo {year} {2013})},\ \Eprint
  {http://arxiv.org/abs/1303.2098} {arXiv:1303.2098 [hep-ph]} \BibitemShut
  {NoStop}%
\bibitem [{\citenamefont {Becattini}\ \emph {et~al.}(2014)\citenamefont
  {Becattini}, \citenamefont {Grossi}, \citenamefont {Bleicher}, \citenamefont
  {Steinheimer},\ and\ \citenamefont {Stock}}]{Becattini:2014hla}%
  \BibitemOpen
  \bibfield  {author} {\bibinfo {author} {\bibfnamefont {F.}~\bibnamefont
  {Becattini}}, \bibinfo {author} {\bibfnamefont {E.}~\bibnamefont {Grossi}},
  \bibinfo {author} {\bibfnamefont {M.}~\bibnamefont {Bleicher}}, \bibinfo
  {author} {\bibfnamefont {J.}~\bibnamefont {Steinheimer}}, \ and\ \bibinfo
  {author} {\bibfnamefont {R.}~\bibnamefont {Stock}},\ }\href {\doibase
  10.1103/PhysRevC.90.054907} {\bibfield  {journal} {\bibinfo  {journal} {Phys.
  Rev. C}\ }\textbf {\bibinfo {volume} {90}},\ \bibinfo {pages} {054907}
  (\bibinfo {year} {2014})},\ \Eprint {http://arxiv.org/abs/1405.0710}
  {arXiv:1405.0710 [nucl-th]} \BibitemShut {NoStop}%
\end{thebibliography}%

\end{document}